\documentclass[journal,draftclsnofoot,onecolumn,12pt]{IEEEtran}
\usepackage{amsthm,amssymb,graphicx,multirow,amsmath,color,amsfonts}
\usepackage[update,prepend]{epstopdf}
\usepackage[noadjust]{cite}
\usepackage[latin1]{inputenc}
\usepackage{tikz}
\usepackage{bbm} 
\usepackage{pdfpages}
\usepackage{tabulary}
\usepackage{multirow}
\usepackage{comment}
\usepackage{subfigure}
\usepackage{float}
\usepackage{algorithmic}
\usepackage{graphicx}
\usepackage{textcomp}
\usepackage{xcolor}
\usepackage{epstopdf}
\def\BibTeX{{\rm B\kern-.05em{\sc i\kern-.025em b}\kern-.08em
	T\kern-.1667em\lower.7ex\hbox{E}\kern-.125emX}}

\def\nb0{{\mathbf{0}}}
\def\nb1{{\mathbf{1}}}

\newtheorem{lemma}{Lemma}

\newtheorem{definition}{Definition}

\newtheorem{theorem}{Theorem}

\newtheorem{remark}{Remark}

\allowdisplaybreaks 
\begin{document}
\title{Energy Efficiency Analysis of Charging
	Pads-powered UAV-enabled Wireless Networks\\}

\date{\today}
\author{
	Yujie Qin, Mustafa A. Kishk, {\em Member, IEEE}, and Mohamed-Slim Alouini, {\em Fellow, IEEE}
	\thanks{Yujie Qin, Mustafa A. Kishk, and Mohamed-Slim Alouini are with Computer, Electrical and Mathematical Sciences and Engineering (CEMSE) Division, King Abdullah University of Science and Technology (KAUST), Thuwal, 23955-6900, Saudi Arabia (e-mail: yujie.qin@kaust.edu.sa; mustafa.kishk@kaust.edu.sa; slim.alouini@kaust.edu.sa).} 
	
}

\maketitle
\begin{abstract}
	This paper analyzes the energy efficiency of a novel system model where unmanned aerial vehicles (UAVs) are used to provide coverage for user hotspots (user clusters) and are deployed on charging pads to enhance the flight time. We introduce a new notion of "cluster pairs" to capture the dynamic nature of the users spatial distribution in order to exploit one of the top advantages of UAVs, which is the mobility and relocation flexibility. Using tools from stochastic geometry, we first derive a new distance distribution that is vital for the energy efficiency analysis. Next, we compute the coverage probability under two deployment strategies: (i) one UAV per cluster pair, and (ii) one UAV per cluster. Finally, we compute the energy efficiency for both strategies. Our numerical results reveal which of the two strategies is better for different system parameters. Our work investigates some new aspects of the UAV-enabled communication system such as the dynamic density of users and the advantages or disadvantages of one- or two-UAV deployment strategies per cluster pair. By considering the relationships between the densities of user cluster pairs and the charging pads, it is shown that an optimal cluster pair density exists to maximize the energy efficiency.
\end{abstract}

\section{Introduction}
As a result of rapid technological progress, unmanned aerial vehicles (UAVs, also known as drones) play an increasingly essential role in enhancing the performance of current and future wireless networks \cite{8316776,li2018uav,zeng2016wireless}. The primary benefits of UAV-enabled networks include the improvement of the coverage probability by establishing line-of-sight (LoS) links with ground users  \cite{fotouhi2019survey}, \cite{8833522}, deployment in dangerous environments or in natural disasters, such as severe snow storms or earthquakes, and more services available to remote Internet of Things (IoT) users in rural areas \cite{cheng2019space}.

In addition, unlike the fixed terrestrial base stations (TBSs), UAVs can easily satisfy the dynamic traffic demands by functioning as aerial BSs, owing to their flexibility, mobility, and 3D-deployment ability \cite{mozaffari2019tutorial}. At places where the spatial distributions of active users continuously changing over time, UAVs are more suitable and eligible as  they are flexible enough to optimize their locations in real time. Under the circumstance in which the users' locations exhibit certain degree of spatial clustering and dynamic variants, UAVs can help offload TBSs by offering an additional dynamic capacity \cite{9205314}. For instance, in some scenarios where the densities of active users vary with time rapidly (e.g., commercial area and office area), UAVs will travel to the user clusters with higher user densities and, therefore, supply the users with stable connectivity, making UAVs, a feasible and practical alternative.

Generally, one of the most popular tools for analyzing cellular networks is stochastic geometry \cite{9420290}, and the most widely used cluster model is the Poisson cluster process (PCP)~\cite{7809177}, such as Matern cluster process (MCP) and Thomas cluster process (TCP)~\cite{7792210}. However, there is always an underlying assumption of homogeneity, which assumes the user density keeps a constant value and does not capture the dynamic variants of the system. Generally, analysis becomes more complex when considering the dynamic variants of the system due to the correlation on time and locations. This paper develops a novel framework that takes the dynamic changes of user densities in a day into account and studies the system's dynamic performance based on stochastic geometry. 


\subsection{Related Work}
Literature has many studies related to this work which can be categorized into: (i) optimization of energy efficiency of a cellular network, (ii) energy efficiency of UAV-enabled wireless network, and (iii) stochastic geometry-based analysis of UAV-enabled wireless network. For each of these areas, a brief introduction is discussed in the following lines.

{\em Optimization of energy efficiency of a cellular network.} Authors in \cite{peng2015fronthaul} comprehensively surveyed recent advances in system architectures and key techniques related to the impact of the constrained fronthaul on spectral efficiency, energy efficiency, and open issues on software-defined networking. Based on stochastic geometry and the deployment of sleeping strategies, authors in \cite{soh2013energy} studied the associated trade-offs of energy-efficient cellular networks. They derived the success probability and formulated the minimization of power consumption problem and maximization of energy efficiency problem. Authors in \cite{6575091} analyzed the minimization of the network energy cost by optimizing the BS density for both homogeneous and heterogeneous cellular networks. The results revealed  which type of BSs should be deployed or slept first when the traffic load is low. In \cite{6567876}, the authors focused on how small cell access points can be used to offload the traffic and how they can enhance energy efficiency. They also evaluated a distributed sleep-mode strategy and the trade-off between traffic offloading and energy consumption based on stochastic geometry.  

{\em Energy efficiency of UAV-enabled wireless network.}
A comprehensive survey of UAV-enabled networks was given in \cite{8843851}, which included in-depth discussions on UAV-based communication systems, classification based on application, altitude, and network, and summarized some open issues and future research directions. Authors in \cite{8941314} studied the energy efficiency of UAV-assisted network with wireless charging. While UAVs aimed to serve multiple users,  energy harvesting time, transmission power, mission complete time and UAV trajectories were jointly considered to minimize the total energy consumption.	Authors in \cite{7572068} and \cite{zeng2017energy} considered a single fixed-wing UAV behaving as a relay between two ground terminals and maximized the system throughput by optimizing the transmit power and trajectory of the UAV. In \cite{7974285}, the authors proposed a novel method to increase the spectral efficiency of UAV small cell by autonomous reposition according to the mobility and activities of users. A self-organized approach with full three-dimensional spatial flexibility was studied in \cite{7565073} to provide the highest possible average spectral energy efficiency in a disaster scenario, subject to interference from the remaining UAVs. In \cite{6965778} a non-stationary fixed-wing UAV-based relay was analyzed, the authors proposed to maximize the energy efficiency by optimizing the circular maneuvering and communication scheme. In \cite{7279205}, BS sleeping strategy was proposed to maximize the energy efficiency of a two-tier downlink cellular network.

{\em Stochastic geometry-based analysis of UAV-enabled wireless network.}
A tutorial on the stochastic geometry-based analysis of cellular networks was provided in \cite{elsawy2016modeling} and \cite{6524460}, which briefly introduced point processes, characterized the interference, and exact error performance. In addition, they presented the outage and ergodic rate analysis and showed that stochastic geometry is capable of capturing realistic network performance. A heterogeneous network composed of terrestrial and aerial BSs was considered in \cite{8833522} and \cite{galkin2019stochastic} where the locations were modeled as two independent Poisson point processes (PPPs). For this setup, the authors first derived accurate expressions of coverage probability and average data rate and then obtained the approximated equations by using the upper bound of the Gamma distribution. Author in \cite{9444343,9153823,qin2021drone} studied the availability probability and revised coverage probability of UAVs subject to the limited capacity of charging stations, where the locations of UAVs and charging stations were modeled as two independent PPPs, and users were modeled as MCP. In \cite{9271904,9091128,9205314}, authors studied a system with tethered UAV, which denotes a UAV connected to a ground station to get a stable power supply and a reliable data link but with restricted mobility.

While the existing literature focus on the spatial distribution of UAVs and TBSs, and some strategies such as sleep-mode to optimize energy efficiency, there is no work to capture the influence of the dynamic variation of user spatial density on UAV deployment and system energy efficiency.
\subsection{Contribution}
This paper investigates the energy efficiency of a UAV-enabled wireless network while capturing the dynamic changes in the user spatial densities. In addition, to avoid the flight time limitation of UAVs, we consider a scenario in which the UAV deploys itself on top of a carefully chosen charging pad. To achieve that objective, we introduce user cluster pairs where the user density in the clusters in each pair vary with time. Modeling the locations of the user cluster pairs as a Poisson bipolar network and the locations of the UAV charging pads as an independent PPP, we investigate the performance of two UAV deployment approaches. The contributions of this paper can be summarized as follows:

{\em Novel Framework and Performance Metrics:} Since the user densities are different according to the time of a day, we introduce a novel framework, the user cluster pair, which consists of two user clusters with varying densities of the user. In this way, we are able to analyze the influence of the dynamic user densities on the system performance. We first define the switching frequency as the number of changes of the user density per day. We then compare the energy efficiency under different switching frequencies and two types of UAV deployment, one UAV deployed for each cluster pair and one UAV deployed for each user cluster, respectively. 

{\em Evaluation of Traveling Distance:} Using tools from stochastic geometry, we derive the expression for the traveling distance, which is defined as the distance between two charging pads that are the nearest to two user clusters in a user cluster pair. The resulting expression and its approximation match well with the simulation and are readily used in the performance analyses.

{\em Energy Efficiency:} Given the aforementioned expression of traveling distance, we are able to study the influence of dynamic user density on the system performance. We first derive the expressions for coverage probability as functions of traveling distance. Next, we obtain the energy efficiency of the above two scenarios, respectively, under the approximation of traveling distance distribution.

{\em System-Level Insights:} By comparing the energy efficiency of the aforementioned two scenarios, our numerical results reveal various useful system insights. We show that user densities do have a significate impact on system performance. For instance, deploying two UAVs is not always better than only one UAV deployed in some scenarios, owing to the interference and relationships between the densities of users. More details are provided in Section \ref{sec_numerical}.

\section{System Model}
\label{sec_sys_mod}
In wireless networks, the locations of users tend to be clustered and the density of users in a cluster varies with time. For example, users are typically concentrated in an office area in the morning or commercial area at night. These two user clusters exist at the same time, but the user densities vary with the time of a day. For instance, the user density of the office cluster is higher than the commercial cluster in the morning and lower at night. Such two user clusters behave like cluster pairs, while the density of users of one cluster is higher than the other one but their relationship changes with time. In our system, we assume that each user cluster pair consists of two user clusters, say $x_m$ and $x_n$, and they are located at a fixed distance $d_{\rm nm}$ in a random orientation. In this case, the user densities of $x_m$ and $x_n$ form the density pairs ($\lambda_{\rm m,h}$,$\lambda_{\rm n,l}$) and ($\lambda_{\rm m,l}$,$\lambda_{\rm n,h}$), where the subscript $h$ and $l$ denote the higher and lower densities of users. That is, $\lambda_{\rm m,h} > \lambda_{\rm n,l}$ and $\lambda_{\rm n,h} > \lambda_{\rm m,l}$, and we are interested in analyzing the influence of switching frequency on the system performance. Here, switching frequency denotes the number of changing of density pairs between ($\lambda_{\rm m,h}$,$\lambda_{\rm n,l}$) and ($\lambda_{\rm m,l}$,$\lambda_{\rm n,h}$) per day.

Different from the previous work \cite{9444343}, we use charging pads in this work to prolong the service time of UAVs. By doing so, UAVs can provide service without being interrupted. Assume that the charging pads are located on the top of buildings at a fixed altitude $h$.  Particularly, we consider that these buildings are able to connect to the network via fiber cables, hence, no need for UAVs to wirelessly backhaul.  UAVs are deployed on charging pads to get a stable power supply and provide service to users in user cluster to offload TBSs.  Note that deploying UAVs on the charging pads is different from roof-top BSs, which is flexible, has low maintenance and installation cost and high energy efficiency.

\begin{figure}[ht]
	\centering
	\includegraphics[width=0.7\columnwidth]{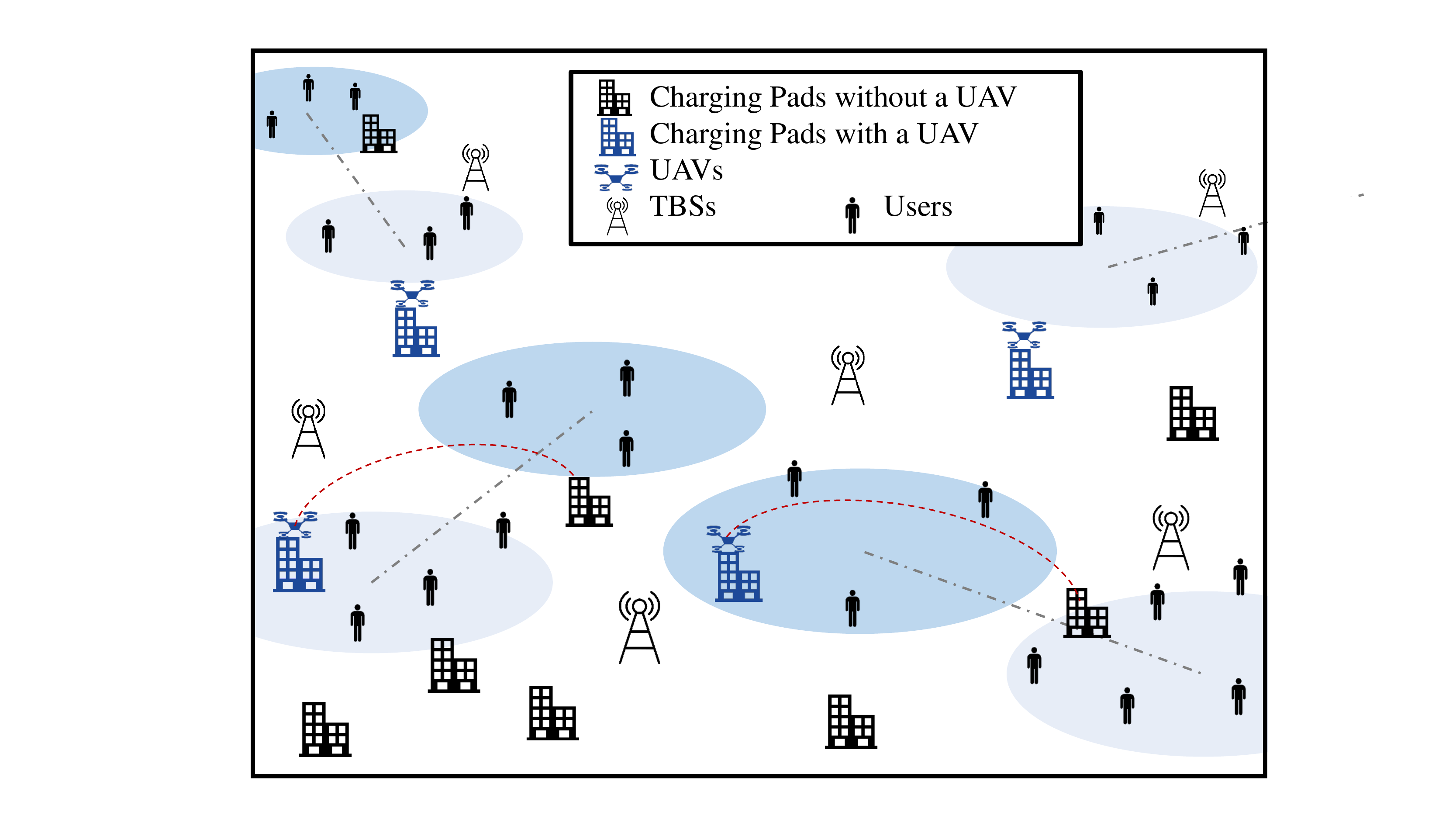}
	\caption{System model.}
	\label{sys_1}
\end{figure}

We consider a dynamic multi-access cellular network that consists of the aforementioned user cluster pairs, UAVs, charging pads, and TBSs, where the switching frequency of user cluster pairs is $n_t$. The locations of charging pads and TBSs are modeled as two independent PPPs, $\Phi_{\rm c}$ and $\Phi_{\rm t}$, with densities $\lambda_{\rm c}$ and $\lambda_{\rm t}$, respectively. The locations of user cluster pairs are modeled as a Poisson bipolar network model $\Phi_{\rm user}$ with density $\lambda_{\rm user}$, as shown in Fig.~\ref{sys_1}. The locations of UAVs are conditioned on user cluster pairs and denoted by $\Phi_{\rm u}$. The locations of  users are uniformly distributed in  user clusters with radii $r_c$, modeled as MCP, and the densities of users are $\lambda_{\rm \{m,n\},\{h,l\}}$ as defined. The locations and densities of UAVs will be explained in details in the following part of this paper.

\begin{table}[ht]\caption{Table of Notations}
	\centering
	\begin{center}
		\resizebox{1\columnwidth}{!}{
			\renewcommand{\arraystretch}{1}
			\begin{tabular}{ {c} | {c} }
				\hline
				\hline
				\textbf{Notation} & \textbf{Description} \\ \hline
				$\Phi_{\rm c}$, $\Phi_{\rm t}$; $\lambda_{\rm c}$, $\lambda_{\rm t}$ & PPP of charging pads, PPP of TBSs; density of the charging pads, density of TBSs\\ \hline
				$x_m$, $x_n$, $d_{\rm nm}$ & The reference user cluster pair, distance between centers of $x_m$ and $x_n$\\ \hline
				$\lambda_{\rm m,\{h,l\}}$, $\lambda_{\rm n,\{h,l\}}$ & User densities of $x_m$ and $x_n$ at different times, respectively\\ \hline
				$\Phi_{\rm user}$,$\Phi_{u}$ & Poisson bipolar network of user cluster pairs, set of locations of UAVs\\\hline
				$\lambda_{\rm user}$, $\lambda_{\rm u,1}$, $\lambda_{\rm u,2}$ &  Densities of user cluster pairs, UAVs of the frist and the second scenarios, respectively\\ \hline
				$C_m$, $C_n$; $L$ & The nearest charging pads to the user clusters $x_m$ and $x_n$, respectively; distance between $C_m$ and $C_n$ \\ \hline
				$R_{\rm mm}$, $R_{\rm mn}$; $\theta_1$,  $\theta_3$ & Distances between the $C_m$ and the cluster center of $x_m$ and $x_n$, respectively; related angles \\ \hline
				$R_{\rm um}$, $R_{\rm un}$ & Distances between $C_m$ and the reference user in $x_m$ and $x_n$, respectively \\ \hline
				$R_{\rm t}$, $R_{\rm nn}$, $\theta_2$ & Distance between the reference user and its nearest TBS, $C_n$ and cluster center of $x_n$, and related angle\\ \hline
				$n_t$, $b_w$, $\gamma$ &   Switching frequency of the cluster UAV, bandwidth, SINR threshold
				\\ \hline
				$p_{\rm s}$, $p_{\rm m}$; $p_{\rm tba}$ & Service-, travel-related power consumption of UAVs; total power consumption of TBSs\\\hline
				$\mathcal{A}_{\rm LoS}$, $\mathcal{A}_{\rm NLoS}$, $\mathcal{A}_{\rm TBS}$& Probability of associating with LoS/NLoS UAV, event of associating with the nearest TBS\\\hline
				${\rm EE}$, ${\rm SE}$, $P_{\rm tot}$ & Energy efficiency,  spectrum efficiency, total energy consumption\\
				\hline\hline
		\end{tabular}}
	\end{center}
\end{table}


We compare the energy efficiency of UAV deployments in two scenarios: one UAV for each cluster pair, or two UAVs for each cluster pair.  Without loss of generality, we perform our analysis at a typical user cluster pair in which the center of $x_m$ is located at the origin and the center of $x_n$ is located at $(d_{\rm nm},0)$. Let $C_m$ and $C_n$ denote the nearest charging pads to the user cluster $x_m$ and $x_n$, respectively.

\begin{figure}[ht]
	\centering
	\includegraphics[width=0.8\columnwidth]{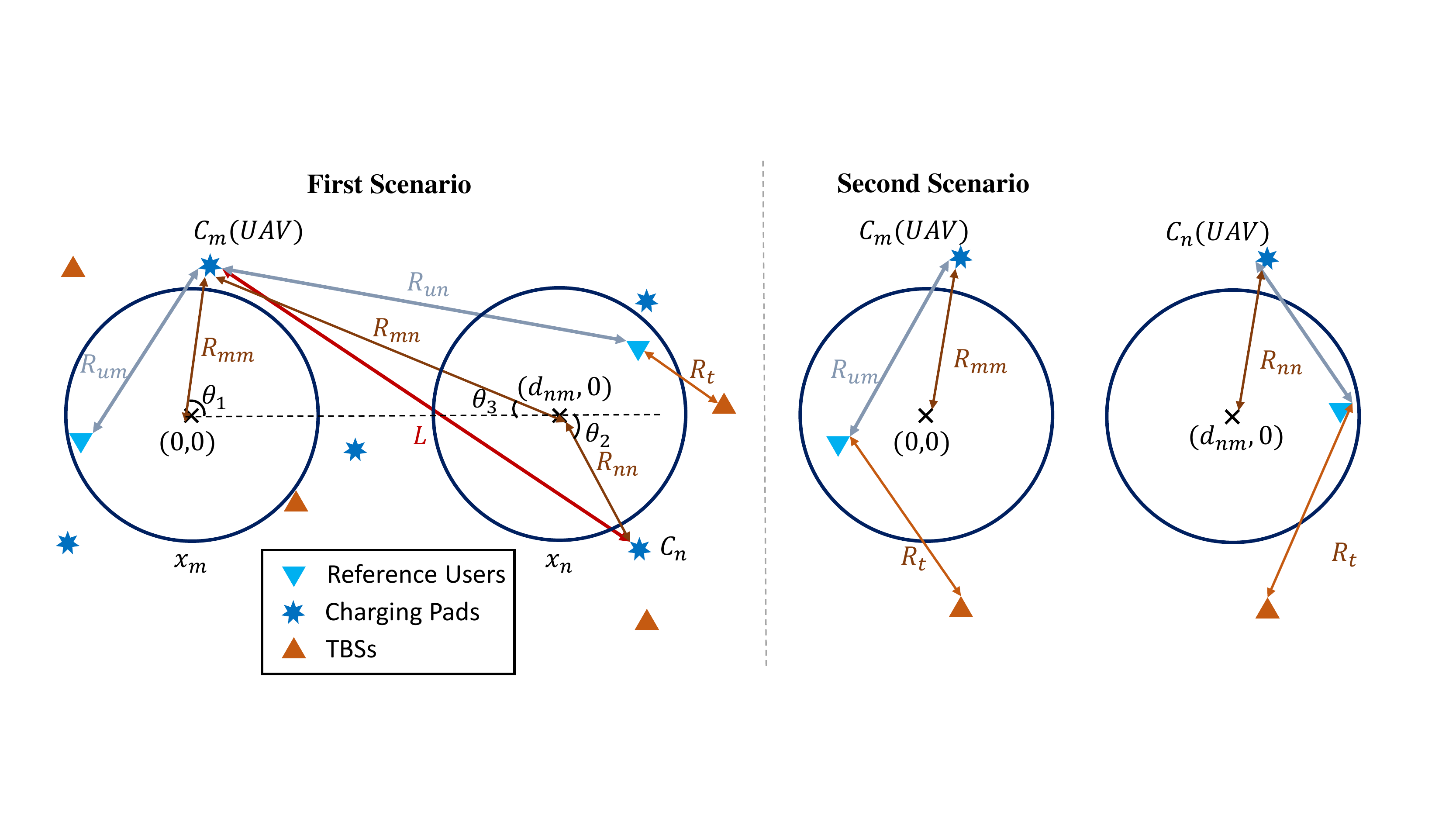}
	\caption{System model of the first scenario (left) and the second scenario (right).}
	\label{Fig_sys_s1}
\end{figure}

\begin{definition}[The First Scenario]
	In the case of the first scenario, each user cluster pair has one UAV deployed at the charging pad which is the nearest to the user cluster that has a higher user density. In other words, UAV locates at $C_m$ when user density pair is $(\lambda_{\rm m,h}, \lambda_{\rm n,l})$ and $C_n$ when  $(\lambda_{\rm m,l}, \lambda_{\rm n,h})$, as shown in Fig.~\ref{Fig_sys_s1}. Therefore, the density of UAVs is equal to the density of user cluster pairs.
\end{definition}
\begin{definition}[The Second Scenario]
	In the case of the second scenario, each user cluster pair has two UAVs deployed, one for $C_m$ and one for $C_n$ if they are different. Otherwise, only one UAV deployed. That is, we only deployed one UAV when the same charging pad is the nearest to both user clusters.
\end{definition}
\begin{remark}
	In the frist scenario, the density of UAVs $\lambda_{\rm u,s_1}$ equals to the density of user cluster pairs $\lambda_{\rm user}$ for sure. However, in the second scenario, the density of UAVs is a function of $d_{\rm nm}$ and $\lambda_{\rm c}$. When $\lambda_{\rm c}$ is high and $d_{\rm nm}$ is large, $\lambda_{\rm u,s_2}$ has a high probability of being equivalent to $2\lambda_{\rm user}$. This is because there will be a very low probability of having the same charging pad being closest to both clusters.
\end{remark}



\subsection{Distance Distribution}
As mentioned in the system model, we assume that the cluster UAV locates at the charging pad which is the nearest to the user cluster that has a higher user density in the first scenario. It implies that the cluster UAV needs to change its location if the density pair of the user clusters switches (e.g., UAV needs to travel from $C_m$ to $C_n$, if the density pair changes from ($\lambda_{\rm m,h}$,$\lambda_{\rm n,l}$) to ($\lambda_{\rm m,l}$,$\lambda_{\rm n,h}$)). While we consider that UAVs cannot provide service when traveling, the traveling time is negligible compared with the time spent on charging pads. 

\begin{definition}[Traveling Distance]\label{def_L}
	Let $L$ be the distance between two charging pads $C_m$ and $C_n$,
	which is conditioned on $R_{\rm mm}$ and $\theta_1$, the distance between $C_m$ and $x_m$, and its relative angle. To simplify the complexity of the analysis, we assume that $R_{\rm mm}$ is shorter than $d_{nm}$.
	
	Let $\mathbb{P}_{\rm (0|R_{\rm mm},\theta_1)}$ denote the probability that $L$ equal to zero in the first scenario. Note that $\mathbb{P}_{\rm (0|R_{\rm mm},\theta_1)}$ is also the probability that only one UAV deployed in the second scenario.
\end{definition}
Therefore, the densities of UAVs in the above two scenarios are, respectively, given by
\begin{align}
	\lambda_{\rm u,1} &= \lambda_{\rm user},\nonumber\\
	\lambda_{\rm u,2} &= (2-\mathbb{E}_{\rm R_{mm},\theta_1}[\mathbb{P}_{\rm (0|R_{\rm mm},\theta_1)}])\lambda_{\rm user},
\end{align}
where $\lambda_{\rm user}$ is the density of user clusters, $\lambda_{\rm u,1}$ and $\lambda_{\rm u,2}$ are the densities of UAVs in the first and the sceond scenario, respectively.

\subsection{User Association}
We assume that users associate with the cluster UAV or the nearest TBS that provides the strongest average received power. Note that the access priority in this work is limited for each UAV to only its own cluster users and this is commonly known as closed access \cite{dhillon2012modeling}.
Without loss of generality, we focus on the reference user randomly selected from the user cluster pair $x_m$ and $x_n$. Let $q_{m,\{h,l\}}$ and $q_{n,\{l,h\}}$ be the probabilities of the reference user being located in the user clusters $x_m$ and $x_n$ at different time of a day, respectively,
\begin{align}
	\label{eq_qmn}
	q_{m,\{h,l\}} &= \frac{\lambda_{\rm m,\{h,l\}}}{\lambda_{\rm m,\{h,l\}}+\lambda_{\rm n,\{l,h\}}},\nonumber\\
	q_{n,\{l,h\}} &= \frac{\lambda_{\rm n,\{l,h\}}}{\lambda_{\rm m,\{h,l\}}+\lambda_{\rm n,\{l,h\}}}.
\end{align}

When the reference user associates with the cluster UAV, the received power is given by,
\begin{align}
	p_{\rm u} = \left\{
	\begin{aligned}
		p_{\rm l} &= \eta_{\rm l}\rho_{\rm u}G_{\rm l}D_{\rm \{um,un\}}^{-\alpha_{\rm l}}, \text{in the case of LoS},\\
		p_{\rm n} &= \eta_{\rm n}\rho_{\rm u}G_{\rm n}D_{\rm \{um,un\}}^{-\alpha_{\rm n}}, \text{in the case of NLoS},\\
	\end{aligned}
	\right.
\end{align}
in which $\rho_{\rm u}$ is the transmit power of UAVs, $\alpha_{\rm l}$ and $\alpha_{\rm n}$ are path-loss exponent, $G_{\rm l}$ and $G_{\rm n}$ present the fading gains that follow gamma distribution with shape and scale parameters ($m_{\rm l},\frac{1}{m_{\rm l}}$) and ($m_{\rm n},\frac{1}{m_{\rm n}}$), $\eta_{\rm l}$ and $\eta_{\rm n}$ denote the mean additional losses for LoS and NLoS transmissions, respectively, $D_{\rm \{um,un\}} = \sqrt{h^2+R_{\rm \{um,un\}}^2}$, where $R_{\rm \{um,un\}}$ are the horizontal between the reference user in $x_m$ or $x_n$ and the cluster UAV, respectively. The occurrence probability of a LoS link established by the reference user and the cluster UAV is given in ~\cite{6863654} as
\begin{align}
	P_{\rm l}(r) & =  \frac{1}{1+a \exp(-b(\frac{180}{\pi}\arctan(\frac{h}{r})-a))} ,
\end{align}
where $a$ and $b$ are two environment variables (e.g., urban, dense urban, and highrise urban), and $h$ is the altitude of the UAV. Consequently, the probability of NLoS link is $P_{\rm n}(r)=1-P_{\rm l}(r)$.

When the user  associates with the nearest TBS, the received power is
\begin{align}
	p_{\rm t} &= \rho_{\rm t} H R_{\rm t}^{-\alpha_{\rm t}},
\end{align}
where $H$ is the fading gain that follows exponential distribution with average power of unity, $\rho_{\rm t}$ is the transmit power of TBSs and $R_{\rm t}$ is the distance between the user and the nearest TBS.

We then define the association probability with UAVs, and an association event that denotes the reference user being served by the nearest TBS.
\begin{definition}[Association Probability and Event]
	Conditioned on the serving base station is at distance $r$ away, let $\mathcal{A}_{\rm NLoS}(r)$ and $\mathcal{A}_{\rm LoS}(r)$ be the probabilities that the reference user associates with the cluster N/LoS UAV, and
	let $\mathcal{A}_{\rm TBS}(r)$ be the event that the user associates with the nearest TBS,
	\begin{align}
		\mathcal{A}_{\rm NLoS}(r) = \mathbb{P}(p_n(r)>p_t),\nonumber\\
		\mathcal{A}_{\rm LoS}(r) = \mathbb{P}(p_l(r)>p_t),\nonumber\\
		\mathcal{A}_{\rm TBS}(r) = \mathbbm{1}(p_t(r)>p_u).
	\end{align}
	The reason of defining $\mathcal{A}_{\rm TBS}(r)$ as an association event will be explained in detail in Remark \ref{remark_ass}.
\end{definition}	

If the signal-to-interference-plus-noise ratio (SINR) of the established channel between the reference user and its serving base station is beyond the designated threshold, this user is said to be successfully served. We define the probability of successfully serving as coverage probability.

\begin{definition}[Coverage Probability]\label{def_cov}
	In the first scenario, the coverage probability $P_{\rm cov,s_1}$ is given by
	\begin{align}
		P_{\rm cov,s_1} =& \alpha P_{\rm cov,s_1|C_m}+(1-\alpha) P_{\rm cov,s_1|C_n} \nonumber\\
		=& \alpha (q_{m,h}P_{\rm cov,s_1|C_m,q_{m,h}}+q_{n,l}P_{\rm cov,s_1|C_m,q_{n,l}})+(1-\alpha) (q_{m,l}P_{\rm cov,s_1|C_n,q_{m,l}}+q_{n,h}P_{\rm cov,s_1|C_n,q_{n,h}}),
	\end{align}
	where $\alpha$ is ratio of the time of cluster UAV spent at $C_m$ and the total time of a day. $P_{\rm cov,s_1|c_{\{m,n\}},q_{m,\{h,l\}}}$ and $P_{\rm cov,s_1|c_{\{m,n\}},q_{n,\{l,h\}}}$ are the coverage probabilities of the reference user in $x_m$ and $x_n$ when UAV locates at $C_m$ and $C_n$ are, respectively, given by
	\begin{align}
		P_{\rm cov,s_1|C_{\{m,n\},qm,\{h,l\}}} = \sum_{b s} \mathbb{E}\left[\mathcal{A}_{b s}(r) \mathbb{P}(\operatorname{SINR} \geq \gamma \mid r, b s)\right],\quad bs\in{\rm \{N/LoS,TBS\}},
	\end{align}
	where $\gamma $ is the SINR threshold.
	
	In the case of the second scenario, the coverage probability  $P_{\rm cov,s_2}$ is given by
	\begin{align}
		P_{\rm cov,s_2} &= P_{\rm cov,s_2,U}+P_{\rm cov,s_2,T},
	\end{align}
	where $P_{\rm cov,s_2,U}$ and $P_{\rm cov,s_2,T}$ are the coverage probabilities of the cluster UAVs and the nearest TBS,
	\begin{align}
		P_{\rm cov,s_2,U} &= \sum_{\rm N/LoS}\mathbb{E}[\mathcal{A}_{\rm N/LoS}(r)\mathbb{P}({\rm SINR}(r)\geq \gamma\mid r, {\rm N/LoS})],\nonumber\\
		P_{\rm cov,s_2,T} &= \mathbb{E}[\mathcal{A}_{\rm TBS}(r)\mathbb{P}({\rm SINR}(r)\geq \gamma\mid r, {\rm T})].
	\end{align}
	
	Let $\Phi_{\rm u_l}$ and $\Phi_{\rm u_n}$ be subsets of $\Phi_{\rm u}$, which denote the locations of LoS UAVs and NLoS UAVs, respectively.  Conditioned on the serving BS $b_s$, the SINR and aggregate interference power are defined as
	\begin{align}
		{\rm SINR} =& \frac{\max(p_u,p_t)}{I+\sigma^2},\nonumber\\
		I =& \sum_{n_i\in\Phi_{\rm u_n}/b_{s}}\eta_{\rm n}\rho_{\rm u}G_{\rm n_i}D_{\rm n_i}^{-\alpha_{\rm n}}+\sum_{l_j\in\Phi_{\rm u_l}/b_{s}}\eta_{\rm l}\rho_{\rm u}G_{\rm l_j}D_{\rm l_j}^{-\alpha_{\rm l}}+\sum_{t_k\in\Phi_{\rm t}/b_s}\rho_{\rm t}H_{\rm t_k}D_{\rm t_k}^{-\alpha_{\rm t}},\label{eq_SINR}
	\end{align} 
	in which $\sigma^2$ is the additive white Gaussian noise (AWGN) power, $D_{\rm n_i}$, $D_{\rm l_j}$ and $D_{\rm t_k}$ are the distances between the reference user and the interfering NLoS, LoS UAVs, and TBSs, respectively.
\end{definition}

\subsection{Energy Efficiency}\label{sec_ee}
We consider the UAV's power consumption composed of travel-related power $p_{\rm m}$ and service-related power $p_{\rm s}$. Based on \cite{8663615},  $p_{\rm s}$ is a constant and ${p}_{\rm m}$ is
\begin{align}
	p_{\rm m}=p_{\rm 0}\left(1+\frac{3v^2}{u_{\rm tip}^{2}}\right)+\frac{p_{\rm i}v_{\rm 0}}{v}+\frac{1}{2}d_{\rm 0}\rho s a_1v^{3},\label{eq_eeUAV}
\end{align}
where $p_{\rm 0}$ and $p_{\rm i}$ present the blade profile power and induced power, $U_{\rm tip}$ is the tip speed of the rotor blade, $v_{\rm 0}$ is the mean rotor induced velocity in hover, $\rho$ is the air density, $a_1$ is the rotor disc area, $d_{\rm 0}$ is fuselage drag ratio, $s$ is rotor solidity, and $v$ is the traveling velocity.


The energy efficiency (EE) performance metric in this paper is defined as the ratio of the network throughput to the total power consumption, and we are interested in investigating the influence of switching frequency $n_t$ on the system performance.

\begin{definition}[Energy Efficiency]\label{def_ee}
	EE of a cellular network is defined in \cite{7279205} as,
	\begin{align}\label{eq_ee_def}
		{\rm EE} = \frac{{\rm SE}}{{\rm P_{\rm tot}}},
	\end{align}
	where ${\rm SE}$ is  spectral efficiency and $P_{\rm tot}$ is the total energy consumption of UAVs and TBSs  \cite{8275034},
	\begin{align}
		{\rm SE}
		&= \text{b}_{\rm w}\log_{2}{(1+\gamma)}(\lambda_{\rm u}P_{\rm cov,U}+\lambda_{\rm t}P_{\rm cov,T}), \\
		P_{\rm tot} &= \lambda_{\rm u}(\beta_{\rm tra}(n_t)p_m+\beta_{\rm ser}(n_t)p_s)+\lambda_{\rm t}p_{\rm tbs},\label{eq_Ptot}
	\end{align}
	in which $\text{b}_{\rm w}$ is the transmission bandwidth, $p_{\rm tbs}$ is the overall power consumption of TBSs, $\beta_{\rm ser}(n_t)$ and $\beta_{\rm tra}(n_t)$ are fractions of serving and traveling time of UAVs, respectively.
\end{definition}

\section{Distance Distribution}\label{sec_distance}
To capture the distribution of traveling distance $L$, which is the most critical  distance in our work, we first derive some prerequisite distance distributions.

In the first scenario, since the user cluster pair is symmetrical,  it is enough to analyze the case where the cluster UAV locates at $C_m$ . With that been said, even though we have two user clusters and two nearest charging pads and cluster UAV needs to travel between them, the system performances (both coverage probability and energy efficiency) of UAV locating at $C_m$ and $C_n$ are similar and influenced by the densities of user cluster $x_m$ and $x_n$ only. Therefore, in the following parts of this paper, we assume that the cluster UAV locates at $C_m$, and all the distances are obtained based on $R_{\rm mm}$ and $\theta_1$, which is the distance between $C_m$ and the center of $x_m$ and related angle.
\begin{lemma}[Distribution of $R_{\rm mm}$ and $R_{\rm nn}$]
	As mentioned in Definition \ref{def_L}, $R_{\rm mm}$ is assumed to be less than $d_{nm}$. Thus, the probability density function (PDF) of $R_{\rm mm}$ is,
	\begin{align}
		f_{\rm R_{\rm mm}}(r) 
		&= \frac{2\pi \lambda_{\rm c} r e^{-\pi \lambda_{\rm c}r^2}}{1-e^{-\pi\lambda_{\rm c}d_{\rm nm}^2}},\quad 0\leq r \leq d_{\rm nm}.
	\end{align}
	Let $R_{\rm nn}$ be the distance between $C_n$ and the center of $x_n$,
	conditioned on $R_{\rm mm}$ and $\theta_1$, the PDF of $R_{\rm nn}$ is
	\begin{small}
		\begin{align}
			&f_{\rm R_{\rm nn}}(r|R_{\rm mm},\theta_1)= \left\{
			\begin{aligned}
				&2\lambda_{\rm c}r\pi\exp(-\lambda_{\rm c}\pi r^2), \quad\text{\rm if}\quad 0<r < d_{nm}-R_{\rm mm},\\
				&2\lambda_{\rm c}r(\pi-\theta_{d}(r))\exp\bigg(-\lambda_{\rm c}\bigg(\pi r^2-\int_{d_{nm}-R_{\rm mm}}^{r}2\theta_{d}(z)z{\rm d}z\bigg)\bigg),\\
				&\text{\rm if}\quad d_{\rm nm}-R_{\rm mm}<r<R_{\rm mn},\\
			\end{aligned}\right.
		\end{align}
	\end{small}
	where $\theta_d$ is a function of $d_{nm}$ and $R_{\rm mm}$,
	\begin{align}
		\theta_{d}(r,R_{\rm mm}) &= \arccos(\frac{d_{\rm nm}^2+r^2-R_{\rm mm}^2}{2d_{\rm nm}r}).
	\end{align}
\end{lemma}

We then derive the  distributions of the distances between the cluster UAV and the reference user in $x_m$ and $x_n$, $R_{\rm um}$ and $R_{\rm un}$, respectively, conditioned on $R_{\rm mm}$.
\begin{lemma}[Distribution of $R_{\rm um}$ and $R_{\rm un}$]
	In the first scenario, let $R_{\rm um}$ and $R_{\rm un}$ be the horizontal distances between the cluster UAV and the reference user in user clusters $x_m$ and $x_n$, respectively. Conditioned on $R_{\rm \{mm,mn\}}<r_c$, the PDF of $R_{\rm um}$ and $R_{\rm un}$ are given by,
	\begin{align}
		&f_{\rm R_{\{um,un\}}}(r|R_{\rm mm})=\left\{ 
		\begin{aligned}
			&\frac{2r}{r_c^2},\quad \text{\rm if} \quad 0<r<r_c-R_{\rm \{mm,mn\}},\\
			&\frac{2r}{\pi r_c^2}\arccos\bigg(\frac{r^2-r_c^2+R_{\rm \{mm,mn\}}^2}{2R_{\rm \{mm,mn\}}r}\bigg), \text{\rm if}\quad r_c-R_{\rm \{mm,mn\}}<r<R_{\rm \{mm,mn\}}+r_c,\\
		\end{aligned} \right.
	\end{align}
	otherwise, if $R_{\rm \{mm,mn\}}>r_c$, we have
	\begin{align}
		f_{\rm R_{\rm\{um,un\}}}(r|R_{\rm mm})&=
		\frac{2r}{\pi r_c^2}\arccos(\frac{r^2-r_c^2+R_{\rm \{mm,mn\}}^2}{2R_{\rm \{mm,mn\}}r}),\quad\text{\rm if} \quad R_{\rm \{mm,mn\}}-r_c<r<R_{\rm \{mm,mn\}}+r_c,
	\end{align}
	where the subscript $\{um,un\}$ corresponding to $\{mm,mn\}$, respectively,
	and $R_{\rm mn}$ is the distance between $C_m$ and the center of $x_n$, which can be represented as function of $R_{\rm mm}$ and $\theta_1$
	\begin{align}
		R_{\rm mn} &= \sqrt{R_{\rm mm}^2+d_{\rm nm}^2-2d_{\rm nm}R_{\rm mm}\cos(\theta_1)}.
	\end{align}
\end{lemma}
Now that we have obtained all the required distance distributions, we can derive the distribution of traveling distance $L$.
\begin{theorem}[Distribution of $L$]
	We first compute the probability of $L=0$, denoted by $\mathbb{P}_{\rm (0|R_{\rm mm},\theta_1)}$,
	\begin{align}
		\mathbb{P}_{\rm (0|R_{\rm mm},\theta_1)} = \exp(-\lambda_{\rm c}({\rm Area}(R_{\rm mm},\theta_1)), \label{l_0}
	\end{align}
	in which,
	\begin{align}
		{\rm Area}(R_{\rm mm},\theta_1) =& R_{\rm mm}^2(\pi-\theta_1)+R_{\rm mn}^2(\pi-\theta_3(R_{\rm mm},\theta_1))+R_{\rm mm} d_{\rm nm} \sin(\theta_1),\nonumber\\
		\theta_3(R_{\rm mm},\theta_1) =& \arccos(\frac{d_{\rm nm}^2+R_{\rm mn}(R_{\rm mm},\theta_1)^2-R_{\rm mm}^2}{2d_{\rm nm}R_{\rm mn}(R_{\rm mm},\theta_1)}).
	\end{align}
	Conditioned on $R_{\rm mm}$ and $\theta_1$, the CDF of L is given in $F_{\rm L}(l)$, please refer to Appendix \ref{app_eq_dis_L} for more detail.
	\begin{IEEEproof}
		{See Appendix \ref{proof_dis_L}.}
	\end{IEEEproof}
\end{theorem}
\begin{remark}\label{rem_L_app}
	Note that the above distribution of L is conditioned on $R_{\rm mm}$ and $\theta_1$, and we need to take the expectation over $R_{\rm mm}$, which is a random variable. Hence, computing the unconditioned distribution of $L$ is very complex and takes a long time. Therefore, in the following paper, we use the expectation of $R_{\rm mm}$ to  approximate the distribution of $L$. More numerical detail will be explained in Section \ref{sec_numerical}.
\end{remark}

\section{Coverage Probability}
To capture the energy efficiency of the aforementioned two scenarios, we need to obtain the coverage probabilities first. In the following theorem, we aim to characterize the association probability and event. Recall the association policy in Sec. \ref{sec_sys_mod}, the reference user associates with the cluster UAV or the nearest TBS, which provides the strongest average received power.
\begin{theorem}[Association Probability and Event]\label{lem_ass}	
	Recall that $\mathcal{A}_{\rm NLoS}(r)$ and $\mathcal{A}_{\rm LoS}(r)$ are the probabilities that the user associates with the cluster N/LoS UAV.
	Conditioned on the cluster UAV is at distance $r$ away with channel type N/LoS, the association probabilities are given as,
	\begin{align}
		\mathcal{A}_{\rm LoS}(r) &= \exp(-\pi \lambda_{\rm t}d_{\rm lt}^{2}(r)),\nonumber\\
		\mathcal{A}_{\rm NLoS}(r) &= \exp(-\pi \lambda_{\rm t}d_{\rm nt}^{2}(r)),
	\end{align}
	in which,
	\begin{align}
		d_{\rm lt}(r) &= (\rho_{\rm t}/\rho_{\rm u})^{\frac{1}{\alpha_{\rm t}}}\eta_{\rm l}^{-\frac{1}{\alpha_{\rm t}}}r^{\frac{\alpha_{\rm l}}{\alpha_{\rm t}}},\nonumber\\
		d_{\rm nt}(r) &= (\rho_{\rm t}/\rho_{\rm u})^{\frac{1}{\alpha_{\rm t}}}\eta_{\rm n}^{-\frac{1}{\alpha_{\rm t}}}r^{\frac{\alpha_{\rm n}}{\alpha_{\rm t}}}.
	\end{align}
	
	Recall $\mathcal{A}_{\rm TBS-N/LoS,\{m,n\}}(r)$ is the event that the user associates with the TBS. Conditioned on the serving TBS is a distance $r$ away and the UAV has the channel type N/LoS at distance $R_{\rm \{um,un\}}$ away,
	\begin{align}
		&\mathcal{A}_{\rm TBS-LoS,\{m,n\}}(r,R_{\rm \{um,un\}})  =  \mathbbm{1}\bigg({\rm LoS(R_{\rm \{um,un\}})}\bigg)\mathbbm{1}\bigg(R_{\rm \{um,un\}}>\sqrt{d_{\rm tl}^2(r)-h^2}\bigg),\nonumber\\
		&\mathcal{A}_{\rm TBS-NLoS,\{m,n\}}(r,R_{\rm \{um,un\}})  = \mathbbm{1}\bigg({\rm NLoS(R_{\rm \{um,un\}})}\bigg)\mathbbm{1}\bigg(R_{\rm \{um,un\}}>\sqrt{d_{\rm tn}^2(r)-h^2}\bigg),
	\end{align}
	where the subscript $\{m,n\}$ denote the location of the reference user, in $x_m$ or $x_n$ and corresponding distance $R_{\rm um}$ or $R_{\rm un}$, respectively,
	\begin{align}
		d_{\rm tl}(r) &= \max\bigg((\rho_{\rm u}/\rho_{\rm t})^{\frac{1}{\alpha_{\rm l}}}\eta_{\rm l}^{\frac{1}{\alpha_{\rm l}}}r^{\frac{\alpha_{\rm t}}{\alpha_{\rm l}}},h\bigg), \nonumber\\
		d_{\rm tn}(r) &= \max\bigg((\rho_{\rm u}/\rho_{\rm t})^{\frac{1}{\alpha_{\rm n}}}\eta_{\rm n}^{\frac{1}{\alpha_{\rm n}}}r^{\frac{\alpha_{\rm t}}{\alpha_{\rm n}}},h\bigg).
	\end{align}
	\begin{IEEEproof}
		See Appendix \ref{app_ass}.
	\end{IEEEproof}
\end{theorem}
\begin{remark}\label{remark_ass}
	Note that in Theorem \ref{lem_ass}, $\mathcal{A}_{\rm NLoS}(r)$ and $\mathcal{A}_{\rm LoS}(r)$ are the association probabilities but $\mathcal{A}_{\rm TBS-N/LoS,\{m,n\}}(r)$ is an association event. The reason is that the association with a TBS is conditioned on $R_{\rm \{um,un\}}$, and should take the expectation over $R_{\rm \{um,un\}}$. However, the Laplace transform of the interference is also conditioned on $R_{\rm \{um,un\}}$. Hence, we write this event as indicator function and take the expectation over $R_{\rm \{um,un\}}$ together. The Laplace transform of the interference will be introduced in the following lemma.
\end{remark}
Laplace transform of the aggregate interference is the final requirement to derive the coverage probability, which is provided in the following lemma.
\begin{lemma}[Laplace Transform]\label{lem_laplace}
	Conditioned on associating with the cluster UAV $bs$ at distance $r$, the  Laplace transform of the aggregate interference power is given by
	\begin{align}
		\mathcal{L}_{\rm I_u,\{s_1,s_2\}}(s,r) =& \exp\bigg\{-2\pi\lambda_{\rm u,\{1,2\}}\biggl(\int_{0}^{\infty}\bigg[1-\bigg(\frac{m_{\rm n}}{m_{\rm n}+s\eta_{\rm n}\rho_{\rm u}(x^2+h^2)^{-\frac{\alpha_{\rm n}}{2}}}\bigg)^{m_{\rm n}}\bigg]xP_{\rm n}(x){\rm d}x\nonumber\\
		&\quad+\int_{0}^{\infty}\bigg[1-\bigg(\frac{m_{\rm l}}{m_{\rm l}+s\eta_{\rm l}\rho_{\rm u}(x^2+h^2)^{-\frac{\alpha_{\rm l}}{2}}}\bigg)^{m_{\rm l}}\bigg]xP_{\rm l}(x){\rm d}x\biggl)\bigg\}\nonumber\\
		&\times \exp\biggl(-2\pi\lambda_{\rm t}\int_{t(r)}^{\infty}\bigg[1-\bigg(\frac{1}{1+s\rho_{\rm t}x^{-\alpha_{\rm t}}}\bigg)\bigg]x{\rm d}x\biggl),
	\end{align}
	where the subscript $\{s_1,s_2\}$ denote the first or the second scenario,
	\begin{align}\label{eq_tr}
		t(r)=\left\{ 
		\begin{aligned}
			d_{\rm nt}(r),  & \quad \text{\rm if} \quad bs \in \Phi_{\rm U_n},\\
			d_{\rm lt}(r),  & \quad \text{\rm if} \quad bs \in \Phi_{\rm U_{l}}.\\
		\end{aligned} \right.
	\end{align}
	
	If the reference user associates with the nearest TBS $bs$, which is at distance $r$, conditioned on the hotspot UAV having a LoS or NLoS link with the user, the Laplace transforms of the aggregate interference are, respectively, given by
	\begin{align}
		\mathcal{L}_{\rm I_{\rm tl,\{m,n\}},\{s_1,s_2\}}(s,r,R_{\rm \{um,un\}}) =&\exp\bigg\{-2\pi\lambda_{\rm u,\{1,2\}}\biggl(\int_{0}^{\infty}\bigg[1-\bigg(\frac{m_{\rm n}}{m_{\rm n}+s\eta_{\rm n}\rho_{\rm u}(x^2+h^2)^{-\frac{\alpha_{\rm n}}{2}}}\bigg)^{m_{\rm n}}\bigg]\nonumber\\
		&xP_{\rm n}(x){\rm d}x+\int_{0}^{\infty}\bigg[1-\bigg(\frac{m_{\rm l}}{m_{\rm l}+s\eta_{\rm l}\rho_{\rm u}(x^2+h^2)^{-\frac{\alpha_{\rm l}}{2}}}\bigg)^{m_{\rm l}}\bigg]xP_{\rm l}(x){\rm d}x\biggr)\bigg\}\nonumber\\
		&\times \exp\biggl(-2\pi\lambda_{\rm t}\int_{r}^{\infty}\bigg[1-\bigg(\frac{1}{1+s\rho_{\rm t}x^{-\alpha_{\rm t}}}\bigg)\bigg]x{\rm d}x\biggl)\nonumber\\
		&\times\bigg(\frac{m_{\rm l}}{m_{\rm l}+s\eta_{\rm l}\rho_{\rm u}(R_{\rm \{um,un\}}^2+h^2)^{-\frac{\alpha_{\rm l}}{2}}}\bigg)^{m_{\rm l}},\nonumber\\
		\mathcal{L}_{\rm I_{\rm tn,\{m,n\}},\{s_1,s_2\}}(s,r,R_{\rm \{um,un\}}) =&\exp\bigg\{-2\pi\lambda_{\rm u,\{1,2\}}\biggl(\int_{0}^{\infty}\bigg[1-\bigg(\frac{m_{\rm n}}{m_{\rm n}+s\eta_{\rm n}\rho_{\rm u}(x^2+h^2)^{-\frac{\alpha_{\rm n}}{2}}}\bigg)^{m_{\rm n}}\bigg]\nonumber\\
		&xP_{\rm n}(x){\rm d}x+\int_{0}^{\infty}\bigg[1-\bigg(\frac{m_{\rm l}}{m_{\rm l}+s\eta_{\rm l}\rho_{\rm u}(x^2+h^2)^{-\frac{\alpha_{\rm l}}{2}}}\bigg)^{m_{\rm l}}\bigg]xP_{\rm l}(x){\rm d}x\biggr)\bigg\}\nonumber\\
		&\times \exp\biggl(-2\pi\lambda_{\rm t}\int_{r}^{\infty}\bigg[1-\bigg(\frac{1}{1+s\rho_{\rm t}x^{-\alpha_{\rm t}}}\bigg)\bigg]x{\rm d}x\biggl)\nonumber\\
		&\times\bigg(\frac{m_{\rm n}}{m_{\rm n}+s\eta_{\rm n}\rho_{\rm u}(R_{\rm \{um,un\}}^2+h^2)^{-\frac{\alpha_{\rm l}}{2}}}\bigg)^{m_{\rm n}}.
	\end{align}
	Note that the subscripts $\{s_{\rm 1},s_{\rm 2}\}$ denote the first and the second scenario, where the densities of UAVs are different, $\lambda_{\rm u,1}$ and $\lambda_{\rm u,2}$, respectively,
	\begin{IEEEproof}
		The Laplace transform is computed using MGF of Gamma distribution, PGFL of inhomogeneous PPP, similar to \cite{9444343} Lemma 7.
	\end{IEEEproof}
\end{lemma}

Having developed all the expressions for the relevant distances and the association probabilities, we study the coverage probability as explained in Definition \ref{def_cov}.
\begin{theorem}[Coverage Probability of the First Scenario]\label{theo_cov}
	Recall that $q_{m,\{h,l\}}$ and $q_{n,\{l,h\}}$ are fractions of user densities and denote the probabilities of the reference user being located in the user cluster $x_m$ and $x_n$ at different times.
	Let $P_{\rm cov,s_1|C_m}$ be the coverage probability when the cluster UAV is located at $C_m$,
	\begin{align}
		&P_{\rm cov,s_1|C_m}= q_{m,h}P_{\rm cov,s_1|C_m,q_{m,h}}+q_{n,l}P_{\rm cov,s_1|C_m,q_{n,l}}\nonumber\\
		&= q_{m,h}(P_{\rm cov,s_1,U|C_m,q_{m,h}}+P_{\rm cov,s_1,T|C_m,q_{m,h}})+q_{n,l}(P_{\rm cov,s_1,U|C_m,q_{n,l}}+P_{\rm cov,s_1,T|C_m,q_{n,l}}),
	\end{align}
	in which, $P_{\rm cov,s_1,U|C_m,q_{m,h}}$ and $P_{\rm cov,s_1,U|C_m,q_{n,l}}$ are the conditional coverage probabilities when the reference user locates at $x_m$ or $x_n$, respectively,
	\begin{align}
		P_{\rm cov,s_1,U|C_m,q_{m,h}} &=  P_{\rm cov,s_1,U_l|C_m,q_{m,h}}+P_{\rm cov,s_1,U_n|C_m,q_{m,h}},\nonumber\\
		P_{\rm cov,s_1,U|C_m,q_{n,l}} &=  P_{\rm cov,s_1,U_l|C_m,q_{n,l}}+P_{\rm cov,s_1,U_n|C_m,q_{n,l}},
	\end{align}
	where,
	\begin{align}
		P_{\rm cov,s_1,U_l|C_m,q_{m,h}} &= \int_{h}^{\sqrt{(d_{\rm nm}+r_c)^2+h^2}}P_{\rm l}(\sqrt{r^2-h^2})\mathcal{A}_{\rm LoS}(r)\sum_{k=0}^{m_{\rm l}-1}\frac{(m_{\rm l}g_{\rm l}(r))^k}{k!}\nonumber\\
		&\quad\times(-1)^k \frac{\partial^k}{\partial s^k}\mathcal{L}_{\rm \sigma^2+I_{\rm u,s_1}}(s,r)|_{\rm s=m_{\rm l}g_{\rm l}(r)}f_{\rm R_{\rm um}}(\sqrt{r^2-h^2}|R_{\rm mm}){\rm d}r,\label{eq_s1_cov_ul_um}\\
		P_{\rm cov,s_1,U_n|C_m,q_{m,h}} &= \int_{h}^{\sqrt{(d_{\rm nm}+r_c)^2+h^2}}P_{\rm n}(\sqrt{r^2-h^2})\mathcal{A}_{\rm NLoS}(r)\sum_{k=0}^{m_{\rm n}-1}\frac{(m_{\rm n}g_{\rm n}(r))^k}{k!}\nonumber\\
		&\quad\times(-1)^k \frac{\partial^k}{\partial s^k}\mathcal{L}_{\rm \sigma^2+I_{\rm u,s_1}}(s,r)|_{\rm s=m_{\rm n}g_{\rm n}(r)}f_{\rm R_{\rm um}}(\sqrt{r^2-h^2}|R_{\rm mm}){\rm d}r,\label{eq_s1_cov_un_um}\\
		P_{\rm cov,s_1,U_l|C_m,q_{n,l}} &= \int_{h}^{\sqrt{(2d_{\rm nm}+r_c)^2+h^2}}P_{\rm l}(\sqrt{r^2-h^2})\mathcal{A}_{\rm LoS}(r)\sum_{k=0}^{m_{\rm l}-1}\frac{(m_{\rm l}g_{\rm l}(r))^k}{k!}\nonumber\\
		&\quad\times(-1)^k \frac{\partial^k}{\partial s^k}\mathcal{L}_{\rm \sigma^2+I_{\rm u,s_1}}(s,r)|_{\rm s=m_{\rm l}g_{\rm l}(r)}f_{\rm R_{\rm un}}(\sqrt{r^2-h^2}|R_{\rm mm}){\rm d}r,\label{eq_s1_cov_ul_un}\\
		P_{\rm cov,s_1,U_n|C_m,q_{n,l}}&= \int_{h}^{\sqrt{(2d_{\rm nm}+r_c)^2+h^2}}P_{\rm n}(\sqrt{r^2-h^2})\mathcal{A}_{\rm NLoS}(r)\sum_{k=0}^{m_{\rm n}-1}\frac{(m_{\rm n}g_{\rm n}(r))^k}{k!}\nonumber\\
		&\quad\times(-1)^k \frac{\partial^k}{\partial s^k}\mathcal{L}_{\rm \sigma^2+I_{\rm u,s_1}}(s,r)|_{\rm s=m_{\rm n}g_{\rm n}(r)}f_{\rm R_{\rm un}}(\sqrt{r^2-h^2}|R_{\rm mm}){\rm d}r\label{eq_s1_cov_un_un},
	\end{align}
	in which,
	\begin{align}
		g_{\rm l}(r)=\gamma(\rho_{\rm u}\eta_{\rm l})^{-1} r^{\alpha_{\rm l}}, \nonumber\\
		g_{\rm n}(r)=\gamma(\rho_{\rm u}\eta_{\rm n})^{-1} r^{\alpha_{\rm n}},
	\end{align}
	and,
	\begin{align}
		P_{\rm cov,s_1,T|C_m,q_{m,h}} 	&= \int_{0}^{\infty}2\pi\lambda_{\rm t}r\exp(-\pi\lambda_{\rm t} r^2)\bigg(\int_{\sqrt{d_{\rm tl}^2(r)-h^2}}^{R_{\rm mm}+r_c}f_{\rm R_{um}}(z|R_{\rm mm})P_{\rm l}(z)\mathcal{L}_{\rm \sigma^2+I_{\rm tl,m,s_1}}(\gamma \rho_{\rm t}^{-1}  r^{\alpha_{\rm t}},r,z){\rm d}z\nonumber\\
		&+\int_{\sqrt{d_{\rm tn}^2(r)-h^2}}^{R_{\rm mm}+r_c}f_{\rm R_{um}}(z|R_{\rm mm})P_{\rm n}(z)\mathcal{L}_{\rm \sigma^2+I_{\rm tn,m,s_1}}(\gamma \rho_{\rm t}^{-1}  r^{\alpha_{\rm t}},r,z){\rm d}z\bigg){\rm d}r,\\
		P_{\rm cov,s_1,T|C_m,q_{n,l}}
		&= \int_{0}^{\infty}2\pi\lambda_{\rm t}r\exp(-\pi\lambda_{\rm t} r^2) \bigg(\int_{\sqrt{d_{\rm tl}^2(r)-h^2}}^{R_{\rm mn}+r_c}f_{\rm R_{un}}(z|R_{\rm mm})P_{\rm l}(z)\mathcal{L}_{\rm \sigma^2+I_{\rm tl,n,s_1}}(\gamma \rho_{\rm t}^{-1}  r^{\alpha_{\rm t}},r,z){\rm d}z\nonumber\\
		&+\int_{\sqrt{d_{\rm tn}^2(r)-h^2}}^{R_{\rm mn}+r_c}f_{\rm R_{un}}(z|R_{\rm mm})P_{\rm n}(z)\mathcal{L}_{\rm \sigma^2+I_{\rm tn,n,s_1}}(\gamma \rho_{\rm t}^{-1}  r^{\alpha_{\rm t}},r,z){\rm d}z\bigg){\rm d}r.
	\end{align}
	\begin{IEEEproof}
		Above equations derived by the fact that  (i) the definition: $\Bar{F}_{\rm G}(g)=\frac{\Gamma_{u}(m,g)}{\Gamma(m)}$, where $\Gamma_{u}(m,g)=\int^{\infty}_{mg}t^{m-1}e^{-t}dt$ is the upper incomplete Gamma function, and (ii) the definition $\frac{\Gamma_{u}(m,g)}{\Gamma(m)}=\exp(-g)\sum^{m-1}_{k=0}\frac{g^{k}}{k!}$, similar to \cite{9444343} Theorem 2.
	\end{IEEEproof}
\end{theorem}
\begin{remark}
	As mentioned in Sec. \ref{sec_distance}, it is enough to analyze the scenario where the cluster UAV locates at $C_m$, and the difference of the coverage probability between the cluster UAV located at $C_m$ and $C_n$ (the difference between $P_{\rm cov,s_1|C_n} $ and $P_{\rm cov,s_1|C_m} $) is the probability of the location of the reference user. That is, $q_{m,l}$ and $q_{n,h}$ are not equal to $q_{n,l}$ and $q_{m,h}$, if the user density pairs are different. Hence, the  coverage probability when the cluster UAV located at $C_n$ is
	\begin{align}
		P_{\rm cov,s_1|C_n}&= q_{m,l}(P_{\rm cov,s_1,U|C_n,q_{m,l}}+P_{\rm cov,s_1,T|C_n,q_{m,l}})+q_{n,h}(P_{\rm cov,s_1,U|C_n,q_{n,h}}+P_{\rm cov,s_1,T|s_1,C_n,q_{n,h}}),
	\end{align}
	where $P_{\rm cov,s_1,U|C_n,q_{m,l}}$, $P_{\rm cov,s_1,T|C_n,q_{m,l}}$, $P_{\rm cov,s_1,U|C_n,q_{n,h}}$ and $P_{\rm cov,s_1,T|s_1,C_n,q_{n,h}}$ are exactly the same as $P_{\rm cov,s_1,U|C_m,q_{m,h}}$, $P_{\rm cov,s_1,T|C_m,q_{m,h}}$, $P_{\rm cov,s_1,U|C_m,q_{n,l}}$ and $P_{\rm cov,s_1,T|s_1,C_m,q_{n,l}}$, respectively.
\end{remark}

Now we present an approximation for the above equations by using the upper bound of the CDF of the Gamma distribution in the case that the expressions of coverage probability require evaluating higher-order derivatives of the Laplace transform.

\begin{lemma}[Approximated Coverage Probability]\label{lem_cov_app}
	$	P_{\rm cov,U_{\rm \{l,n\}}|C_m,q_{m,\{h,l\}}}$ and $P_{\rm cov,U_{\{l,n\}}|C_m,q_{n,\{l,h\}}}$ can be approximated by using the upper bound of the CDF of the Gamma distribution as,
	\begin{align}
		P_{\rm cov,U_{\rm \{l,n\}}|C_m,q_{m,h}}=& \int_{h}^{\sqrt{(d_{\rm nm}+r_c)^2+h^2}}P_{\rm \{l,n\}}(\sqrt{r^2-h^2})\mathcal{A}_{\rm \{L,NL\}oS}(r)\sum_{k=1}^{m_{\rm \{l,n\}}}\binom{m_{\rm \{l,n\}}}{k}\nonumber\\
		&\quad\times(-1)^{k+1}\bigg(\mathcal{L}_{\rm \sigma^2+I_{\rm u,s_1}}(k\beta_2m_{\rm \{l,n\}}g_{\rm \{l,n\}}(r)),r\bigg)f_{\rm R_{\rm um}}(\sqrt{r^2-h^2}|R_{\rm mm}){\rm d}r,\nonumber\\
		P_{\rm cov,U_{\{l,n\}}|C_m,q_{n,l}} &= \int_{h}^{\sqrt{(2d_{\rm nm}+r_c)^2+h^2}}P_{\rm l}(\sqrt{r^2-h^2})\mathcal{A}_{\rm \{L,NL\}oS}(r)\sum_{k=1}^{m_{\rm \{l,n\}}}\binom{m_{\rm \{l,n\}}}{k}\nonumber\\
		&\quad\times(-1)^{k+1}\bigg(\mathcal{L}_{\rm \sigma^2+I_{\rm u,s_1}}(k\beta_2m_{\rm \{l,n\}}g_{\rm \{l,n\}}(r)),r\bigg)f_{\rm R_{\rm un}}(\sqrt{r^2-h^2}|R_{\rm mm}){\rm d}r,
	\end{align}
	in which $\beta_{2}=(m_{\rm {\{l,n\}}}!)^{-\frac{1}{m_{\rm {\{l,n\}}}}}$.
	\begin{IEEEproof}
		Approximate coverage probability is derived by using the definition of CCDF of Gamma function, upper incomplete Gamma function and Binomial theorem, similar to \cite{9444343} Lemma 8.
	\end{IEEEproof}
\end{lemma}
Following the same way, we obtain the expressions of the coverage probability of the second scenario.
\begin{theorem}[Coverage Probability of the Second Scenario]\label{theo_cov_2}
	The coverage probability of the second scenario is,
	\begin{align}
		P_{\rm cov,s_2} &= P_{\rm cov,s_2,U}+P_{\rm cov,s_2,T}= P_{\rm cov,s_2,U_l}+P_{\rm cov,s_2,U_n}+P_{\rm cov,s_2,T},
	\end{align}
	where,
	\begin{align}
		P_{\rm cov,s_2,U_l}=& \int_{h}^{\sqrt{(d_{\rm nm}+r_c)^2+h^2}}P_{\rm l}(\sqrt{r^2-h^2})\mathcal{A}_{\rm LoS}(r)\sum_{k=0}^{m_{\rm l}-1}\frac{(m_{\rm l}g_{\rm l}(r))^k}{k!}\nonumber\\
		&\quad\times(-1)^k \frac{\partial^k}{\partial s^k}\mathcal{L}_{\rm \sigma^2+I_{u,s_2}}(s)|_{\rm s=m_{\rm l}g_{\rm l}(r)}f_{\rm R_{\rm um}}(\sqrt{r^2-h^2}|R_{\rm mm}){\rm d}r,\label{eq_s2_cov_Ul}\\
		P_{\rm cov,s_2,U_n}=& \int_{h}^{\sqrt{(d_{\rm nm}+r_c)^2+h^2}}P_{\rm n}(\sqrt{r^2-h^2})\mathcal{A}_{\rm NLoS}(r)\sum_{k=0}^{m_{\rm n}-1}\frac{(m_{\rm n}g_{\rm n}(r))^k}{k!}\nonumber\\
		&\quad\times(-1)^k \frac{\partial^k}{\partial s^k}\mathcal{L}_{\rm \sigma^2+I_{u,s_2}}(s)|_{\rm s=m_{\rm n}g_{\rm n}(r)}f_{\rm R_{\rm um}}(\sqrt{r^2-h^2}|R_{\rm mm}){\rm d}r,\label{eq_s2_cov_Un}\\
		P_{\rm cov,s_2,T}=& \int_{0}^{\infty}2\pi\lambda_{\rm t}r\exp(-\pi\lambda_{\rm t} r^2)\bigg(\int_{\sqrt{d_{\rm tl}(r)^2-h^2}}^{d_{\rm nm}+r_c}\mathcal{A}_{\rm TBS-LoS}(r)\mathcal{L}_{\rm \sigma^2+I_{\rm tl,s_2}}(\gamma \rho_{\rm t}^{-1}  r^{\alpha_{\rm t}},r|x)f_{\rm R_{um}}(x){\rm d}x\nonumber\\
		&+\int_{\sqrt{d_{\rm tn}(r)^2-h^2}}^{d_{\rm nm}+r_c}\mathcal{A}_{\rm TBS-NLoS}(r)\mathcal{L}_{\rm \sigma^2+I_{\rm tn,s_2}}(\gamma \rho_{\rm t}^{-1}  r^{\alpha_{\rm t}},r|x)f_{\rm R_{um}}(x){\rm d}x\bigg){\rm d}r.\label{eq_s2_cov_T}
	\end{align}
\end{theorem}
Similar to the Theorem \ref{theo_cov}, we use the upper bound approximation to simlify the high order derivation terms in the Laplace transform in Theorem \ref{theo_cov_2} of the second scenario.
\begin{lemma}[Approximated Coverage Probability]
	By using the upper bound, $P_{\rm cov,s_2,U_{\{l,n\}}}$ can be written as,
	\begin{align}
		P_{\rm cov,s_2,U_{\{l,n\}}}=&  \int_{h}^{\sqrt{(d_{\rm nm}+r_c)^2+h^2}}P_{\rm \{l,n\}}(\sqrt{r^2-h^2})\mathcal{A}_{\rm \{L,NL\}oS}(r)\sum_{k=1}^{m_{\rm \{l,n\}}}\binom{m_{\rm \{l,n\}}}{k}(-1)^{k+1}\nonumber\\
		&\quad\times\bigg(\mathcal{L}_{\rm \sigma^2+I_{u,s_2}}(k\beta_2m_{\rm \{l,n\}}g_{\rm \{l,n\}}(r))\bigg)f_{\rm R_{\rm um}}(\sqrt{r^2-h^2}|R_{\rm mm}){\rm d}r.
	\end{align}	
\end{lemma}
\section{Energy Efficiency}
Finally, we compute the energy efficiency of these two scenarios. We first derive the expressions of SE and total power consumption.
\begin{lemma}[Spectral Efficiency and Total Power Consumption]
	The SE (bit/sec/m$^2$) can be formulated as,
	\begin{align}
		{\rm SE}_{\{s_1,s_2\}} 
		&=\text{b}_{\rm w}\log_{2}{(1+\gamma)}(\lambda_{\rm u,\{1,2\}}P_{\rm cov,\{s_1,s_2\},U}+\lambda_{\rm t}P_{\rm cov,\{s_1,s_2\},T}),\label{eq_PSE}
	\end{align}
	in which $\text{b}_{\rm w}$ is the transmission bandwidth, while $P_{\rm cov,s_2,U}$ and $P_{\rm cov,s_2,T}$ are defined in (\ref{eq_s2_cov_Ul}), (\ref{eq_s2_cov_Un}) and (\ref{eq_s2_cov_T}), $P_{\rm cov,s_1,U}$ and $P_{\rm cov,s_1,T}$ are
	\begin{align}
		P_{\rm cov,s_1,U}=& \alpha(q_{m,h}P_{\rm cov,s_1,U|C_m,q_{m,h}}+q_{n,l}P_{\rm cov,s_1,U|C_m,q_{n,l}})\nonumber\\
		&+(1-\alpha)(q_{m,l}P_{\rm cov,s_1,U|C_n,q_{m,l}}+q_{n,h}P_{\rm cov,s_1,U|C_n,q_{n,h}}),\\
		P_{\rm cov,s_1,T}=& \alpha(q_{m,h}P_{\rm cov,s_1,T|C_m,q_{m,h}}+q_{n,l}P_{\rm cov,s_1,T|C_m,q_{n,l}})\nonumber\\
		&+(1-\alpha)(q_{m,l}P_{\rm cov,s_1,T|C_n,q_{m,l}}+q_{n,h}P_{\rm cov,s_1,T|C_n,q_{n,h}}).
	\end{align}
	
	Total power consumption ${\rm P_{\rm tot}}$ (Watt/m$^2$) can be formulated as,
	\begin{align}
		P_{\rm tot,s_1} &= \lambda_{\rm u,1}\bigg(n_{t}\frac{2L}{24\times 3600v}p_{\rm m}+(1-n_{t}\frac{2L}{24\times 3600v})p_{\rm s}\bigg)+\lambda_{\rm t}p_{\rm tbs}, \label{eq_Ptot_s1}\\
		P_{\rm tot,s_2} &= \lambda_{\rm u,2}p_{\rm s}+\lambda_{\rm t}p_{tbs},\label{eq_Ptot_s2}
	\end{align}
	where $n_t$ is the frequency of traveling per day, $v$ is the velocity when traveling and in numerical results section, we use the optimal $v$ which minimizes the energy consumption, $p_{\rm m}$ and $p_{\rm s}$ are the motion-related and service-related power, respectively.
	\begin{IEEEproof}
		Recall that $\beta_{\rm ser}(n_t)$ and $\beta_{\rm tra}(n_t)$ are time fractions, the ratio of serving, taveling time to total time of a day, given by
		\begin{align}
			\beta_{\rm ser}(n_t) &= n_{t}\frac{2L}{24\times 3600v},\nonumber\\
			\beta_{\rm tra}(n_t) &= 1-\beta_{\rm ser}(n_t). \label{eq_beta}
		\end{align}
		Proof completes by substituting (\ref{eq_beta}) into (\ref{eq_Ptot}).
	\end{IEEEproof}
\end{lemma}
Recall the energy efficiency introduced in Definition \ref{def_ee}, which is the ratio of the network throughput to the total power consumption.
\begin{theorem}[Energy Efficiency]
	EE in both scenarios are, respectively, given by, 
	\begin{align}
		{\rm EE}_1 &= \mathbb{E}_{\rm L}\bigg[\frac{\text{b}_{\rm w}\log_{2}{(1+\gamma)}(\lambda_{\rm u,1}P_{\rm cov,s_1,U}+\lambda_{\rm t}P_{\rm cov,s_1,T})}{ \lambda_{\rm u,1}\bigg(n_{t}\frac{2L}{24\times 3600v}p_{\rm m}+(1-n_{t}\frac{2L}{24\times 3600v})p_{\rm s}\bigg)+\lambda_{\rm t}p_{\rm tbs}}\bigg],\label{eq_ee_1}\\
		{\rm EE}_2 &= \frac{\text{b}_{\rm w}\log_{2}{(1+\gamma)}(\lambda_{\rm u,2}P_{\rm cov,s_1,U}+\lambda_{\rm t}P_{\rm cov,s_1,T})}{\lambda_{\rm u,2}p_{\rm s}+\lambda_{\rm t}p_{tbs}}.
	\end{align}
	where $n_t$ is the switching frequency, which is the number of times the cluster UAV travels back and forth between $C_m$ and $C_n$ per day.
	\begin{IEEEproof}
		The above equations follow by substituting (\ref{eq_PSE}), (\ref{eq_Ptot_s1}) and (\ref{eq_Ptot_s2}) into (\ref{eq_ee_def}).
	\end{IEEEproof}
\end{theorem}
\begin{remark}\label{rem_ee_app}
	Computing (\ref{eq_ee_1}) requires the PDF of $L$, which needs to take the derivative of the CDF of $L$ given in Appendix \ref{app_eq_dis_L}. Here we use the expectation to approximate (\ref{eq_ee_1}), now the energy efficiency of the first scenario is given as
	\begin{align}
		{\rm EE}_1 &= \frac{\text{b}_{\rm w}\log_{2}{(1+\gamma)}(\lambda_{\rm u,1}P_{\rm cov,s_1,U}+\lambda_{\rm t}P_{\rm cov,s_1,T})}{ \lambda_{\rm u,1}\bigg(n_{t}\frac{2\mathbb{E}[L]}{24\times 3600v}p_{\rm m}+(1-n_{t}\frac{2\mathbb{E}[L]}{24\times 3600v})p_{\rm s}\bigg)+\lambda_{\rm t}p_{\rm tbs}}.
	\end{align}
\end{remark}

\section{Numerical Results}\label{sec_numerical}
In this section, we validate our analytical results with simulations and evaluate the impact of various system parameters on the network performance. To simplify the selection of parameters, we set $\lambda_{\rm m,l} = \lambda_{\rm n,l}$ and $\lambda_{\rm m,h} = \lambda_{\rm n,h}$.

Unless stated otherwise, we use the simulation parameters as listed herein Table \ref{par_val}. 
\begin{table}[ht]\caption{Table of Parameters}\label{par_val}
	\centering
	\begin{center}
		\resizebox{0.7\columnwidth}{!}{
			\renewcommand{\arraystretch}{1}
			\begin{tabular}{ {c} | {c} | {c}  }
				\hline
				\hline
				\textbf{Parameter} & \textbf{Symbol} & \textbf{Simulation Value}  \\ \hline
				Density of TBSs & $\lambda_{\rm t}$ & $10^{-6}$ m$^{-2}$ \\ \hline
				Traveling-related power & $P_{\rm m}$ & 161.8 W \\\hline
				Service-related power & $P_{\rm s}$ & 10 W \\\hline
				UAV altitude & $h$ & 60 m\\\hline
				UAV velocity & $v$ & 18.46 m/s \\\hline
				Radius of MCP disk & $r_c$ & 120 m \\\hline
				N/LoS environment variable & $a, b$ & 25.27,0.2 \\\hline
				Transmission power of UAVs and TBSs& $\rho_{\rm u}$, $\rho_{\rm t}$ & 0.4 W, 10 W\\\hline
				Total power comsuption of TBSs & $p_t$& 318 W \\\hline
				SINR threshold & $\gamma$ & 0 dB \\\hline
				Noise power & $\sigma^2 $ & $10^{-9}$ W\\\hline
				N/LoS and active charging station path-loss exponent & $\alpha_{\rm n},\alpha_{\rm l},\alpha_{\rm t}$ & $4,2.1,4$ \\\hline
				N/LoS fading gain & $m_{\rm n},m_{\rm l}$ & $1,3$ \\\hline
				N/LoS additional loss& $\eta_{\rm n},\eta_{\rm l}$ & $-20,0$ dB 
				\\\hline\hline
		\end{tabular}}
	\end{center}
\end{table}

\begin{figure}[ht]
	\centering
	\includegraphics[width=0.6\columnwidth]{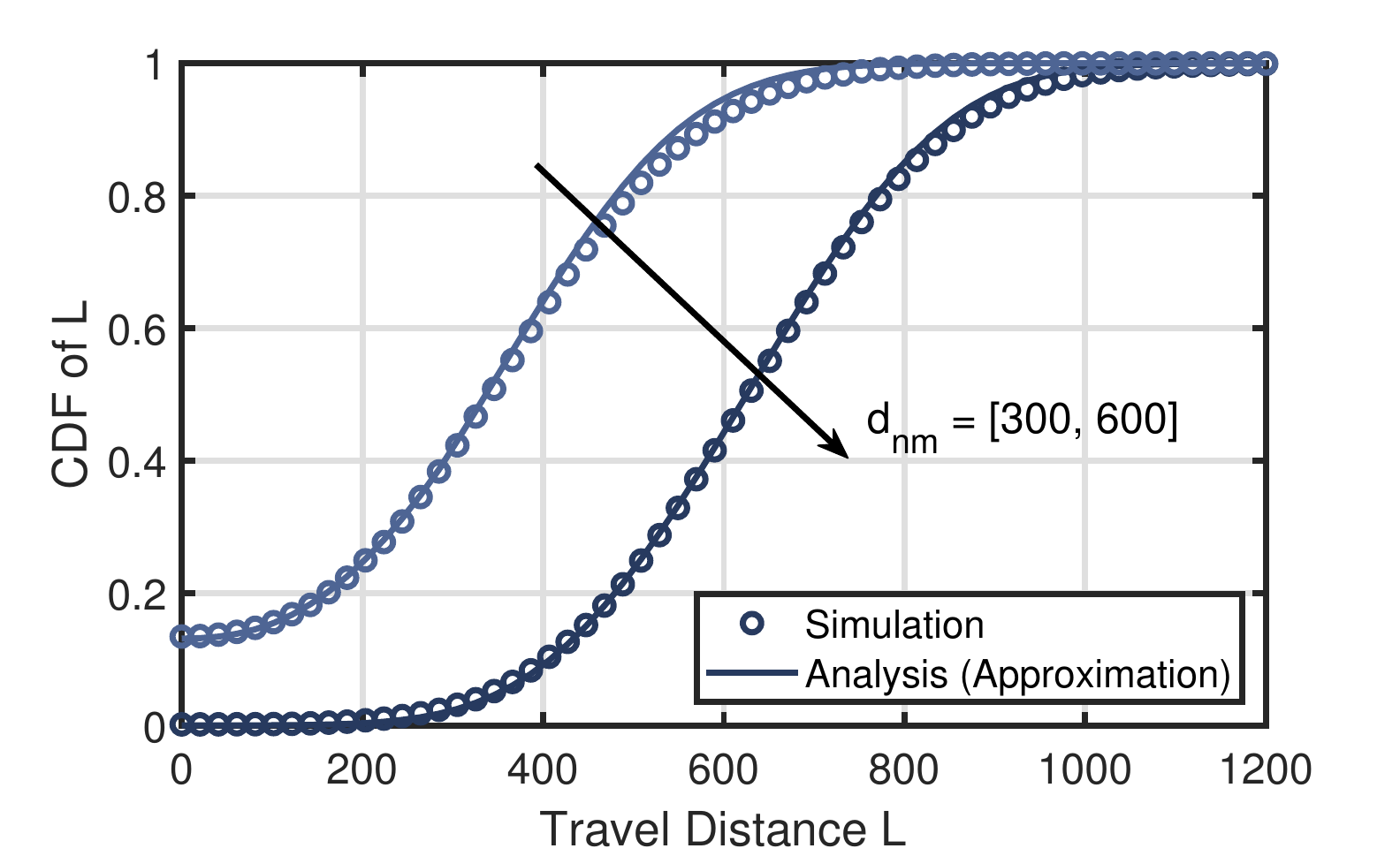}	
	\caption{Using the approximation, the distance distribution of $L$, conditioned on $\lambda_c = 10$ charging pad/km$^2$. }
	\label{fig_dis_L}
\end{figure}

For the simulation of the considered system model, we first compute the accuracy of the approximation of traveling distance $L$, as mentioned in Remark \ref{rem_L_app}. We simulate two cases, $d_{\rm nm}=300$ and $d_{\rm nm}=600$. The traveling distance increases with the increase of $d_{\rm nm}$. In both cases, the results in Fig.~\ref{fig_dis_L} show that using the expectation of $R_{\rm mm}$ provides a close approximation of the distribution of $L$. We then use the approximate distribution of $L$ in the following part of this paper.

\begin{figure}[ht]
	\centering
	\includegraphics[width=0.6\columnwidth]{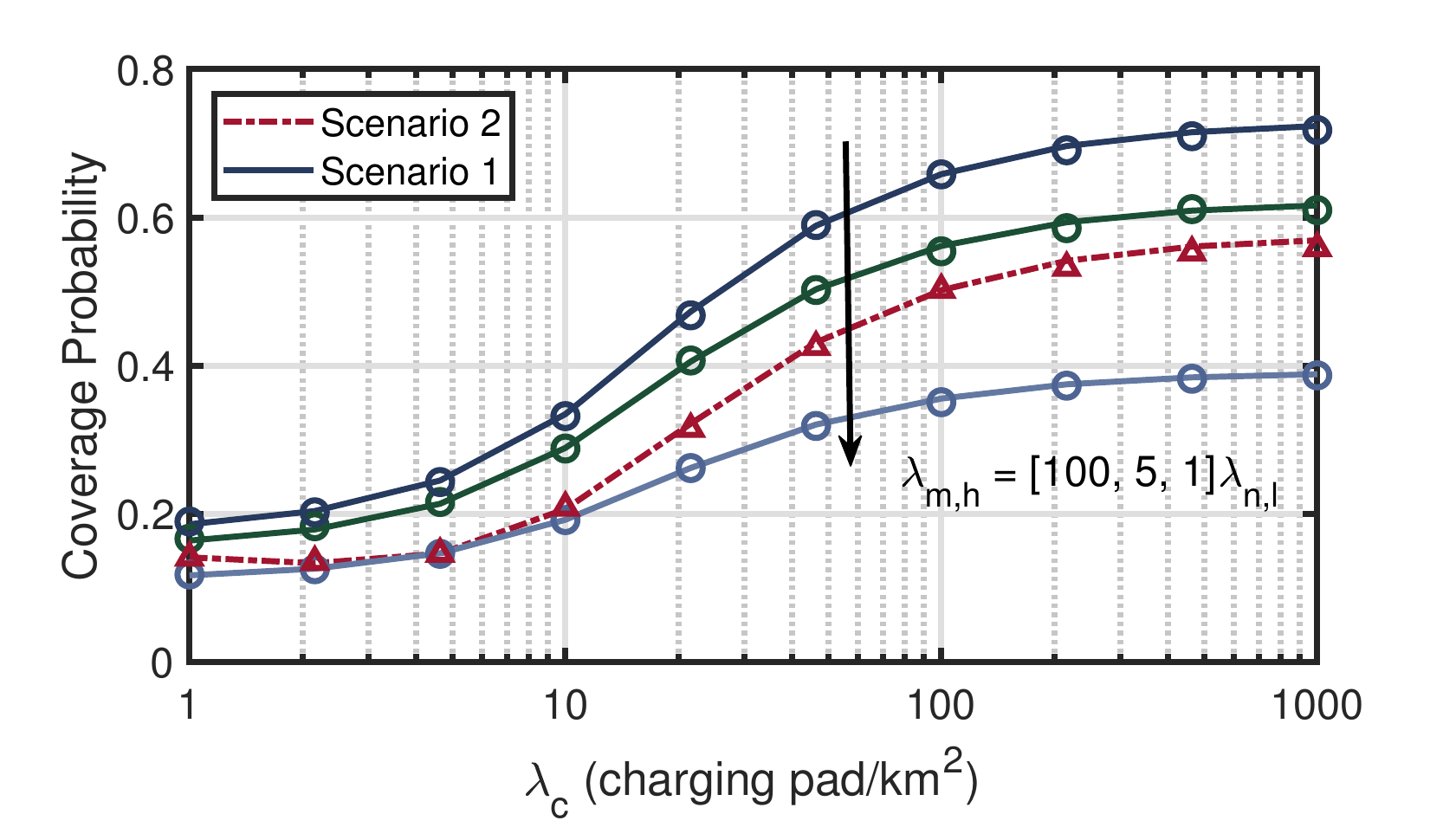}	
	\caption{Coverage probabilities under different densities of charging pads, conditioned on $\lambda_{\rm t} = 1$ TBS/km$^{2}$, $\lambda_{u} = 10$ user cluster pair/km$^{2}$, and $d_{\rm nm} = 300$ m.}
	\label{fig_Pcov_dif_lambda_c}
\end{figure}
\begin{figure}[ht]
	\centering
	\includegraphics[width=0.6\columnwidth]{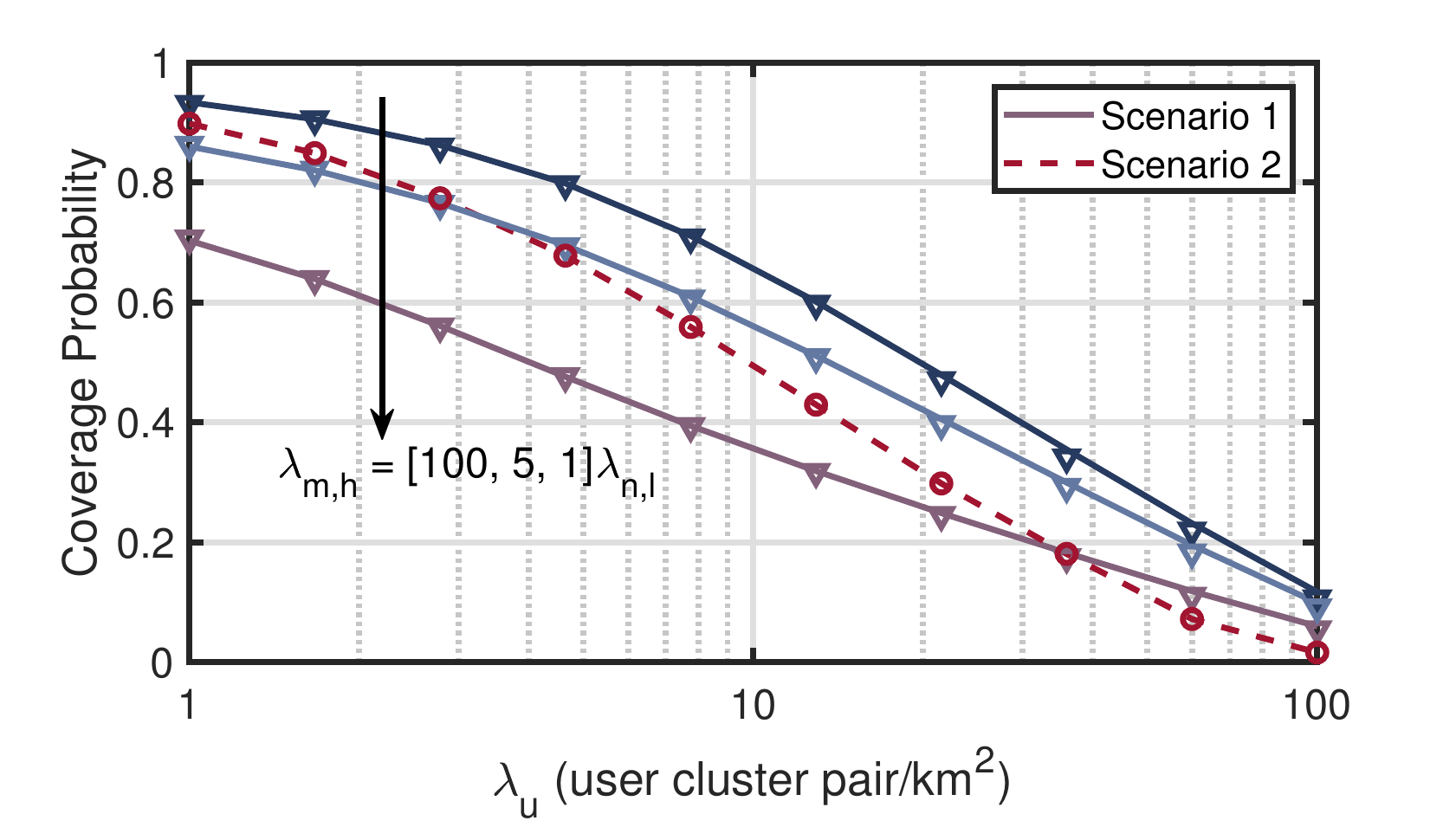}	
	\caption{Coverage probabilities under different densities of user cluster pairs, conditioned on $\lambda_{\rm c} = 100$ charging pad/km$^{2}$, $\lambda_{t} = 1$ TBS/km$^{2}$, and $d_{\rm nm} = 300$ m. }
	\label{fig_Pcov_dif_lambda_u}
\end{figure}

Next, Fig.~\ref{fig_Pcov_dif_lambda_c} and Fig.~\ref{fig_Pcov_dif_lambda_u} show the coverage probabilities increase with the increase of the density of charging pads, and decrease when increasing the density of user cluster pairs.  Both two figures show that the second scenario is not always better than the first one, especially when the density of one user cluster is much higher than that of the other one. The system performance of one UAV per cluster pair is better, if the user density of one user cluster is much higher than another one. The reason is that in the second scenario, even though we have one UAV for each user cluster, and the reference user gets closer to the serving cluster UAV. However, the interference becomes higher due to the increasing density of UAVs compared with the first scenario. In addition, the interference has a higher impact on the coverage probabilities than distance.

\begin{figure}[ht]
	\centering
	\includegraphics[width=0.6\columnwidth]{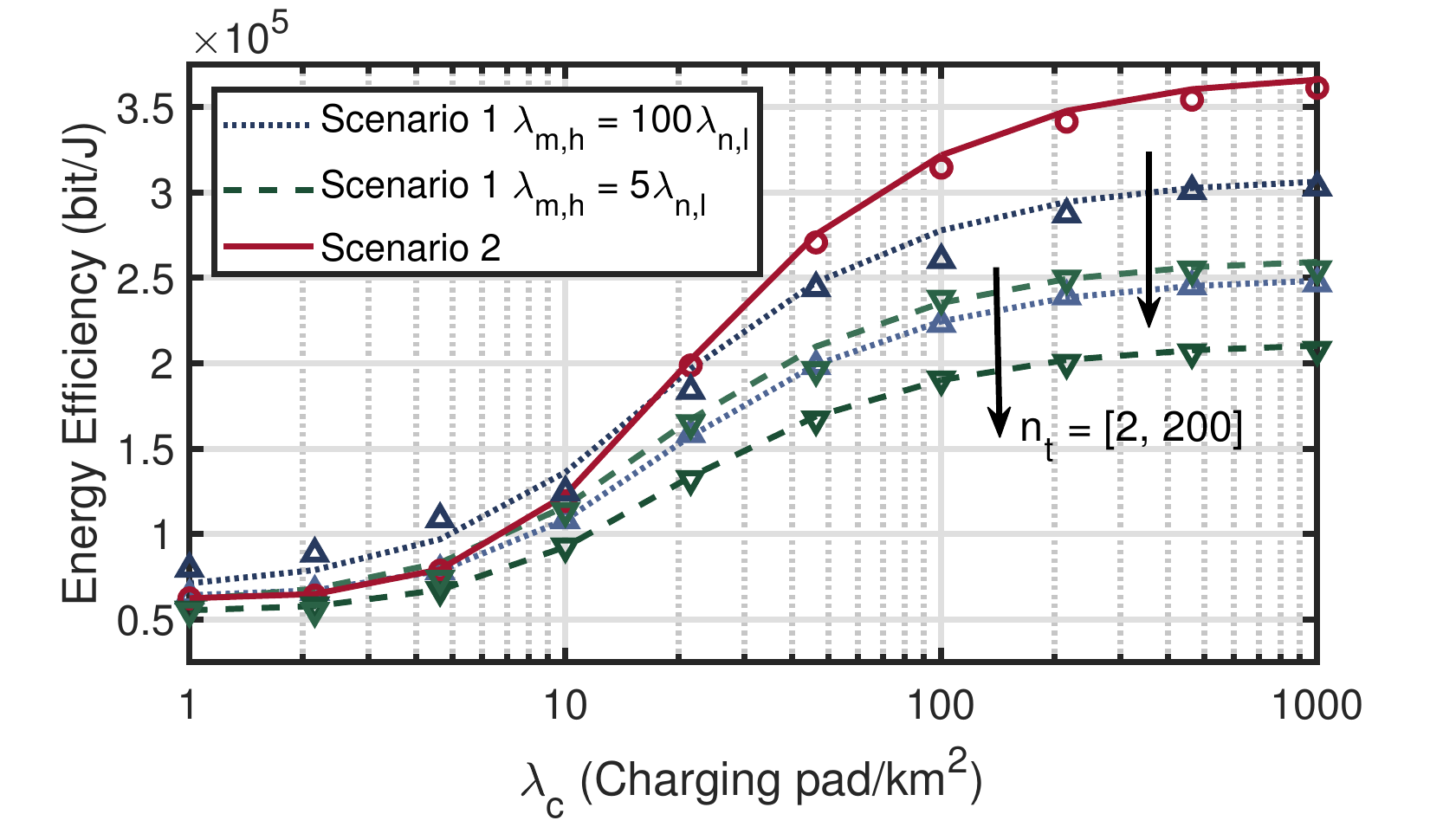}	
	\caption{Energy efficiency under different densities of charging pads, conditioned on $\lambda_{\rm u} = 10$ user cluster pair/km$^{2}$, $\lambda_{t} = 1$ TBS/km$^{2}$, and $d_{\rm nm} = 300$ m. }
	\label{fig_ee_dif_c}
\end{figure}
\begin{figure}[ht]
	\centering
	\includegraphics[width=0.6\columnwidth]{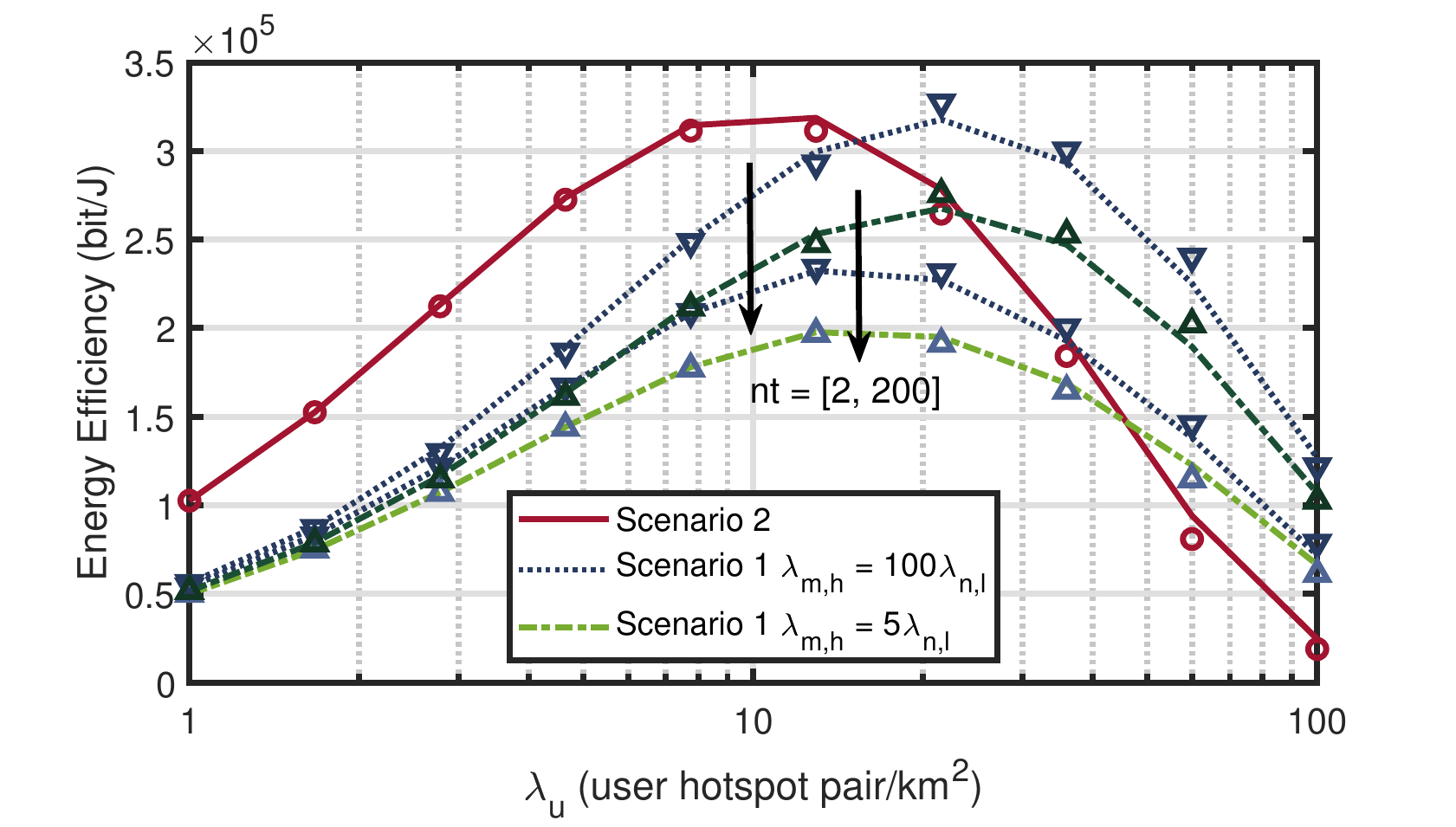}	
	\caption{Energy efficiency under different densities of user cluster pairs, conditioned on $\lambda_{\rm c} = 100$ charging pad/km$^{2}$, $\lambda_{t} = 1$ TBS/km$^{2}$, and $d_{\rm nm} = 300$ m. }
	\label{fig_ee_dif_u}
\end{figure}

Finally, we plot the energy efficiency against $\lambda_{\rm c}$ and $\lambda_{\rm u}$ in Fig.~\ref{fig_ee_dif_c} and Fig.~\ref{fig_ee_dif_u}, respectively. Note that, the gaps between simulation and analysis in energy efficiency when $n_t$ is large ($n_t=200$) are due to the approximation we used in Remark \ref{rem_ee_app}.  As expected, energy efficiency increases with the increase of the densities of charging pads due to the increase of the coverage probability. 	The system's energy efficiency decreases with the increase of $d_{\rm nm}$ since the traveling-related energy consumption is higher than service-related energy. Besides, if two user clusters are too far from each other, we cannot ignore the impact of traveling time on the coverage probability. Fig. \ref{fig_ee_dif_c} shows that the second scenario has a better performance when $\lambda_{\rm c}$ is high, that is because of the higher densities of UAVs. With the increase of the density of charging pads, the probability of $L=0$ decreases dramatically. Hence the densities of UAV, as well as SE, becomes larger compared with the first scenario.

Fig.~\ref{fig_ee_dif_u} shows the influence of $\lambda_{\rm u}$. Energy efficiency increases first due to the increase of coverage probability and densities of UAVs, and then decreases owing to the decreasing of coverage probability because of the higher interference.

The results reveal an interesting system insight that two-UAV deployment strategy does not always achieve a better performance, which also depends on the density of users and
user clusters. For instance, when the density of one user cluster is much greater than the other, the user cluster density is very high or the density of charging pads is low, deploying one UAV per cluster pair is better.
\section{Conclusion}
This paper presented a novel system model that considered the dynamic density of users and captured the energy efficiency in two scenarios, deploying one UAV or two UAVs per cluster pair. Firstly, we derived the PDF of  traveling distance. Next, we computed the coverage probability of the above two scenarios. Finally, we used the aforementioned distance and coverage probability to compute the energy efficiency of the considered system model. We have validated that our distance distribution is accurate and shown that one UAV deployment is better than two UAVs deployment if the density of the user cluster is high or the density of charging pads is low.

Our work investigated some new aspects of the performance of UAV-enabled wireless networks, such as the dynamic variations of user clusters and one or two UAVs deployment, and obtained a new distance distribution. By considering the relationships between the densities of user cluster pairs and charging pads, optimal energy efficiency can be achieved by deploying different numbers of UAVs. A more complicated model can be developed in future work, such as the optimization of UAV locations.
\appendix
\subsection{Distribution of Traveling Distance L}\label{app_eq_dis_L}
We are interested in obtaining the distance distribution of L, which is shown in Fig. \ref{Fig_sys_s1}, the distance between $C_m$ and $C_n$, which are the nearest charging pads to the centers of  user cluster  $x_m$ and $x_n$, respectively. Conditioned on $R_{\rm mm}$ and $\theta_1$, the CDF of $L$ is given by,
\begin{align}
	F_{\rm L}(l) = \mathbb{P}_{\rm (0|R_{\rm mm},\theta_1)}, \quad \text{if} \quad l = 0,
\end{align}
otherwise, 

(i) in the case of $\theta_1+\theta_3(R_{\rm mm},\theta_1) < \pi/2$,
\begin{itemize}
	\item if $0<l\leq R_{\rm mn}+d_{\rm nm}-R_{\rm mm}$,  $F_{\rm L}(l)$ is given by
	$$	\int_{R_{\rm mn}-l}^{R_{2d}}f_3(r,l,R_{\rm mm},\theta_1){\rm d}r+\int_{R_{2d}}^{R_{\rm mn}}f_6(r,l,R_{\rm mm},\theta_1){\rm d}r
	+\mathbb{P}_{\rm (0|R_{\rm mm},\theta_1)}$$
	\item if $R_{\rm mn}+d_{\rm nm}-R_{\rm mm}<l\leq 2R_{\rm mm}\sin(\frac{\theta_1}{2})$,  $F_{\rm L}(l)$ is given by
	$$\int_{R_{\rm mn}-l}^{d_{\rm nm}-R_{\rm mm}}f_3(r,l,R_{\rm mm},\theta_1){\rm d}r+\int_{d_{\rm nm}-R_{\rm mm}}^{R_{2d}}f_4(r,l,R_{\rm mm},\theta_1){\rm d}r+\int_{R_{2d}}^{R_{\rm mn}}f_6(r,l,R_{\rm mm},\theta_1){\rm d}r
	+\mathbb{P}_{\rm (0|R_{\rm mm},\theta_1)}$$
	\item if $2R_{\rm mm}\sin(\frac{\theta_1}{2})\leq l\leq \min(2R_{\rm mm}\sin(\theta_1),R_{\rm mn})$,  $F_{\rm L}(l)$ is given by
	$$	e^{-\lambda_{\rm c}\pi(R_{\rm mn}-l)^2}-\int_{R_{\rm mn}-l}^{d_{\rm nm}-R_{\rm mm}}f_2(r,l,R_{\rm mm},\theta_1){\rm d}r-\int_{d_{\rm nm}-R_{\rm mm}}^{R_{2d}}f_4(r,l,R_{\rm mm},\theta_1){\rm d}r-\int_{R_{2d}}^{R3}f_9(r,l,R_{\rm mm},\theta_1){\rm d}r$$
	\item if $2R_{\rm mm}\sin(\theta_1)\leq l\leq R_{\rm mn}$, $F_{\rm L}(l)$ is given by
	$$\exp(-\lambda_{\rm c}\pi(R_{\rm mn}-l)^2)-\int_{R_{\rm mn}-l}^{d_{\rm nm}-R_{\rm mm}}f_2(r,l,R_{\rm mm},\theta_1){\rm d}r-\int_{d_{\rm nm}-R_{\rm mm}}^{R_{3}}f_4(r,l,R_{\rm mm},\theta_1){\rm d}r$$
	\item if $ R_{\rm mn}< l\leq \min(R_{\rm mn}+d_{\rm nm}-R_{\rm mm},2R_{\rm mm}\sin(\theta_1))$, $F_{\rm L}(l)$ is given by
	$$1-\int_{R_{\rm mn}-l}^{d_{\rm nm}-R_{\rm mm}}f_2(r,l,R_{\rm mm},\theta_1){\rm d}r-\int_{d_{\rm nm}-R_{\rm mm}}^{R_{2d}}f_4(r,l,R_{\rm mm},\theta_1){\rm d}r
	-\int_{R_{2d}}^{R_{\rm mn}}f_9(r,l,R_{\rm mm},\theta_1){\rm d}r$$
	\item if $R_{\rm mn}+d_{\rm nm}-R_{\rm mm}<l\leq 2R_{\rm mm}\sin(\theta_1)$, $F_{\rm L}(l)$ is given by
	$$1-\int_{d_{\rm nm}-R_{\rm mm}}^{R_{2d}}f_4(r,l,R_{\rm mm},\theta_1){\rm d}r
	-\int_{R_{2d}}^{R_{\rm mn}}f_9(r,l,R_{\rm mm},\theta_1){\rm d}r$$
	\item if $ \max(R_{\rm mn},2R_{\rm mm}\sin(\theta_1))\leq l\leq R_{\rm mn}+d_{\rm nm}-R_{\rm mm}$, $F_{\rm L}(l)$ is given by
	$$	1-\int_{R_{\rm mn}-l}^{d_{\rm nm}-R_{\rm mm}}f_2(r,l,R_{\rm mm},\theta_1){\rm d}r-\int_{d_{\rm nm}-R_{\rm mm}}^{R_{\rm mn}}f_4(r,l,R_{\rm mm},\theta_1){\rm d}r$$
	\item if $\max(2R_{\rm mm}\sin(\theta_1),R_{\rm mn}+d_{\rm nm}-R_{\rm mm})<l\leq 2R_{\rm mn}$, $F_{\rm L}(l)$ is given by
	$$1-\int_{l-R_{\rm mn}}^{R_{\rm mn}}f_4(r,l,R_{\rm mm},\theta_1){\rm d}r$$
\end{itemize}
(ii) in the case of $\cos(\pi-\theta_1-\theta_3(R_{\rm mm},\theta_1))>\sin(\theta_1)$,
\begin{itemize}
	\item if $0<l\leq 2R_{\rm mm}\sin(\theta_1)$, $F_{\rm L}(l)$ is given by
	$$\int_{R_{2d}}^{R_{\rm mn}}f_6(r,l,R_{\rm mm},\theta_1){\rm d}r+\mathbb{P}_{\rm (0|R_{\rm mm},\theta_1)}$$
	\item if $2R_{\rm mm}\sin(\theta_1)<l\leq 2R_{\rm mm}\cos(\pi-\theta_1-\theta_3(R_{\rm mm},\theta_1)),$ $F_{\rm L}(l)$ is given by
	$$
	\int_{R_{2d}}^{R_{\rm mn}}f_6(r,l,R_{\rm mm},\theta_1){\rm d}r+\int_{R_{2d}}^{R_{\rm mn}}f_7(r,l,R_{\rm mm},\theta_1){\rm d}r+\mathbb{P}_{\rm (0|R_{\rm mm},\theta_1)}$$
	\item if $2R_{\rm mm}\cos(\pi-\theta_1-\theta_3(R_{\rm mm},\theta_1))<l\leq \min(R_{\rm mn}+d_{\rm nm}-R_{\rm mm}, 2R_{\rm mm}),$ $F_{\rm L}(l)$ is given by
	$$
	\int_{R_{2d}}^{R_{\rm mn}}f_6(r,l,R_{\rm mm},\theta_1){\rm d}r+\int_{R_{2d2}}^{R_{\rm mn}}f_7(r,l,R_{\rm mm},\theta_1){\rm d}r+\int_{l-R_{\rm mn}}^{d_{\rm nm}-R_{\rm mm}}f_1(r,l,R_{\rm mm},\theta_1){\rm d}r$$
	$$+\int_{d_{\rm nm}-R_{\rm mm}}^{R_{2d}}f_5(r,l,R_{\rm mm},\theta_1){\rm d}r+\mathbb{P}_{\rm (0|R_{\rm mm},\theta_1)}$$
	\item if $ R_{\rm mn}+d_{\rm nm}-R_{\rm mm}<l\leq 2R_{\rm mm},$ $F_{\rm L}(l)$ is given by
	$$
	\int_{R_{2d}}^{R_{\rm mn}}f_6(r,l,R_{\rm mm},\theta_1){\rm d}r+\int_{R_{2d2}}^{R_{\rm mn}}f_7(r,l,R_{\rm mm},\theta_1){\rm d}r
	+\int_{l-R_{\rm mn}}^{R_{2d}}f_5(r,l,R_{\rm mm},\theta_1){\rm d}r+\mathbb{P}_{\rm (0|R_{\rm mm},\theta_1)}
	$$
	\item  if $2R_{\rm mm}<l\leq R_{\rm mn},$ $F_{\rm L}(l)$ is given by
	$$
	\exp(-\lambda_{\rm c}\pi(R_{\rm mn}-l)^2)-\int_{R_{\rm mn}-l}^{d_{\rm nm}-R_{\rm mm}}f_2(r,l,R_{\rm mm},\theta_1){\rm d}r-\int_{d_{\rm nm}-R_{\rm mm}}^{R3}f_4(r,l,R_{\rm mm},\theta_1){\rm d}r$$
	\item if $R_{\rm mn}<l\leq R_{\rm mn}+d_{\rm nm}-R_{\rm mm},$ $F_{\rm L}(l)$ is given by
	$$
	1-\int_{l-R_{\rm mn}}^{d_{\rm nm}-R_{\rm mm}}f_2(r,l,R_{\rm mm},\theta_1){\rm d}r-\int_{d_{\rm nm}-R_{\rm mm}}^{R_{\rm mn}}f_4(r,l,R_{\rm mm},\theta_1){\rm d}r$$
	\item if $\max(2R_{\rm mm},R_{\rm mn}+d_{\rm nm}-R_{\rm mm})<l\leq 2R_{\rm mn},$ $F_{\rm L}(l)$ is given by
	$$
	1-\int_{l-R_{\rm mn}}^{R_{\rm mn}}f_4(r,l,R_{\rm mm},\theta_1){\rm d}r$$
\end{itemize}
(iii) in the case of $\theta_1+\theta_3(R_{\rm mm},\theta_1) \geq \pi/2 \quad \text{and} \quad \cos(\pi-\theta_1-\theta_3(R_{\rm mm},\theta_1))\leq\sin(\theta_1)$,
\begin{itemize}
	\item if $0<l\leq 2R_{\rm mm}\cos(\pi-\theta_1-\theta_3)$, $F_{\rm L}(l)$ is given by
	$$
	\int_{R_{2d}}^{R_{\rm mn}}f_6(r,l,R_{\rm mm},\theta_1){\rm d}r+\mathbb{P}_{\rm (0|R_{\rm mm},\theta_1)}$$
	\item if $2R_{\rm mm}\cos(\pi-\theta_1-\theta_3)<l\leq 2R_{\rm mm}\sin(\frac{\theta_1}{2}),$ $F_{\rm L}(l)$ is given by
	$$
	\int_{R_{\rm mn}-l}^{d_{\rm nm}-R_{\rm mm}}f_1(r,l,R_{\rm mm},\theta_1){\rm d}r+\int_{d_{\rm nm}-R_{\rm mm}}^{R_{2d}}f_3(r,l,R_{\rm mm},\theta_1){\rm d}r+\int_{R_{2d}}^{R_{\rm mn}}f_6(r,l,R_{\rm mm},\theta_1){\rm d}r
	+\mathbb{P}_{\rm (0|R_{\rm mm},\theta_1)}$$
	\item $2R_{\rm mm}\sin(\frac{\theta_1}{2})<l\leq \min(2R_{\rm mm}\sin(\theta_1),R_{\rm mn}),$ $F_{\rm L}(l)$ is given by
	$$
	\int_{R_{\rm mn}-l}^{d_{\rm nm}-R_{\rm mm}}f_1(r,l,R_{\rm mm},\theta_1){\rm d}r+\int_{d_{\rm nm}-R_{\rm mm}}^{R_{2d}}f_4(r,l,R_{\rm mm},\theta_1){\rm d}r+\int_{R_{2d}}^{R_{\rm mn}}f_6(r,l,R_{\rm mm},\theta_1){\rm d}r\\
	+\mathbb{P}_{\rm (0|R_{\rm mm},\theta_1)}$$
	\item if $R_{\rm mn}<l\leq \min(2R_{\rm mm}\sin(\theta_1),R_{\rm mn}+d_{\rm nm}-R_{\rm mm}),$ $F_{\rm L}(l)$ is given by
	$$
	1-\int_{l-R_{\rm mn}}^{d_{\rm nm}-R_{\rm mm}}f_2(r,l,R_{\rm mm},\theta_1){\rm d}r-\int_{d_{\rm nm}-R_{\rm mm}}^{R_{2d}}f_4(r,l,R_{\rm mm},\theta_1){\rm d}r-\int_{R_{2d}}^{R_{\rm mn}}f_9(r,l,R_{\rm mm},\theta_1){\rm d}r
	$$
	\item if $R_{\rm mn}+d_{\rm nm}-R_{\rm mm}<l\leq 2R_{\rm mm}\sin(\theta_1),$ $F_{\rm L}(l)$ is given by
	$$
	1-\int_{l-R_{\rm mn}}^{R_{2d}}f_4(r,l,R_{\rm mm},\theta_1){\rm d}r-\int_{R_{2d}}^{R_{\rm mn}}f_9(r,l,R_{\rm mm},\theta_1){\rm d}r$$
	\item if $2R_{\rm mm}\sin(\theta_1)<l\leq R_{\rm mn}$, $F_{\rm L}(l)$ is given by
	$$
	\exp(-\lambda_{\rm c}\pi(R_{\rm mn}-l)^2)-\int_{R_{\rm mn}-l}^{d_{\rm nm}-R_{\rm mm}}f_2(r,l,R_{\rm mm},\theta_1){\rm d}r-\int_{d_{\rm nm}-R_{\rm mm}}^{R3}f_4(r,l,R_{\rm mm},\theta_1){\rm d}r
	+g(l,R_{\rm mm},\theta_1)$$
	\item if $\max(R_{\rm mn}, 2R_{\rm mm}\sin(\theta_1))<l\leq R_{\rm mn}+d_{\rm nm}-R_{\rm mm},$ $F_{\rm L}(l)$ is given by
	$$
	1-\int_{l-R_{\rm mn}}^{d_{\rm nm}-R_{\rm mm}}f_2(r,l,R_{\rm mm},\theta_1){\rm d}r-\int_{d_{\rm nm}-R_{\rm mm}}^{R_{\rm mn}}f_4(r,l,R_{\rm mm},\theta_1){\rm d}r+g(l,R_{\rm mm},\theta_1)$$
	\item if $\max(R_{\rm mn}+d_{\rm nm}-R_{\rm mm}, 2R_{\rm mm}\sin(\theta_1))<l\leq 2R_{\rm mn}$, $F_{\rm L}(l)$ is given by
	$$
	1-\int_{l-R_{\rm mn}}^{R_{\rm mn}}f_4(r,l,R_{\rm mm},\theta_1){\rm d}r+g(l,R_{\rm mm},\theta_1)$$
\end{itemize}
where $g(l,R_{\rm mm},\theta_1) = \int_{R_{2d}}^{R_{2d2}}f_8(r,l,R_{\rm mm},\theta_1){\rm d}r$ if $\theta_1>\frac{\pi}{2}$ and $l<2R_{\rm mm}$, otherwise, $0$,
\begin{table}[ht]\caption{Table of Notations}
	\centering
	\begin{center}
		\resizebox{0.7\columnwidth}{!}{
			\renewcommand{\arraystretch}{1}
			\begin{tabular}{ {c} | {c} }
				\hline
				\hline
				\textbf{Function} & \textbf{Definition} \\ \hline
				$f_1(r,l,R_{\rm mm},\theta_1)$ & $f_{\rm R_{\rm nn}}(r|R_{\rm mm},\theta_1)\frac{\theta(r,l,R_{\rm mm},\theta_1)}{\pi}$\\ \hline
				$f_2(r,l,R_{\rm mm},\theta_1)$ & $f_{\rm R_{\rm nn}}(r|R_{\rm mm},\theta_1)\frac{\pi-\theta(r,l,R_{\rm mm},\theta_1)}{\pi}$\\ \hline
				$f_3(r,l,R_{\rm mm},\theta_1)$ & $f_{\rm R_{\rm nn}}(r|R_{\rm mm},\theta_1)\frac{\theta(r,l,R_{\rm mm},\theta_1)}{\pi-\theta_{d}(r,R_{\rm mm},\theta_1)}$\\ \hline
				$f_4(r,l,R_{\rm mm},\theta_1)$ & $f_{\rm R_{\rm nn}}(r|R_{\rm mm},\theta_1)\frac{\pi-\theta(r,l,R_{\rm mm},\theta_1)}{\pi-\theta_{d}(r,l,R_{\rm mm},\theta_1)}$\\\hline
				$f_5(r,l,R_{\rm mm},\theta_1)$ &  $f_{\rm R_{\rm nn}}(r|R_{\rm mm},\theta_1)\frac{\theta(r,l,R_{\rm mm},\theta_1)-\theta_{d}(r,R_{\rm mm},\theta_1)}{\pi-\theta_{d}(r,R_{\rm mm},\theta_1)}$\\ \hline
				$f_6(r,l,R_{\rm mm},\theta_1)$ & $f_{\rm R_{\rm nn}}(r|R_{\rm mm},\theta_1)\frac{\theta(r,l,R_{\rm mm},\theta_1)+\theta_3(R_{\rm mm},\theta_1)-\theta_{d}(r,R_{\rm mm},\theta_1)}{2\pi-2\theta_{d}(r,R_{\rm mm},\theta_1)}$ \\ \hline
				$f_7(r,l,R_{\rm mm},\theta_1)$ & $f_{\rm R_{\rm nn}}(r|R_{\rm mm},\theta_1)\frac{\theta(r,l,R_{\rm mm},\theta_1)-\theta_3(R_{\rm mm},\theta_1)-\theta_{d}(r,R_{\rm mm},\theta_1)}{2\pi-2\theta_{d}(r,R_{\rm mm},\theta_1)}$ \\ \hline
				$f_8(r,l,R_{\rm mm},\theta_1)$ & $f_{\rm R_{\rm nn}}(r|R_{\rm mm},\theta_1)\frac{-\theta(r,l,R_{\rm mm},\theta_1)+\theta_3(R_{\rm mm},\theta_1)+\theta_{d}(r,R_{\rm mm},\theta_1)}{2\pi-2\theta_{d}}$ \\ \hline
				$f_9(r,l,R_{\rm mm},\theta_1)$ & $f_{\rm R_{\rm nn}}(r|R_{\rm mm},\theta_1)\frac{2\pi-\theta(r,l,R_{\rm mm},\theta_1)-\theta_3(R_{\rm mm},\theta_1)-\theta_{d}(r,R_{\rm mm},\theta_1)}{2\pi-2\theta_{d}(r,R_{\rm mm},\theta_1)}$\\\hline
				$\theta_3(R_{\rm mm},\theta_1)$ & $\arccos(\frac{d_{\rm nm}^2+R_{\rm mn}(R_{\rm mm},\theta_1)^2-R_{\rm mm}^2}{2d_{\rm nm}R_{\rm mn}(R_{\rm mm},\theta_1)})$ \\ \hline
				$\theta(r,l,R_{\rm mm},\theta_1)$ & $\arccos(\frac{R_{\rm mn}(R_{\rm mm},\theta_1)^2+r^2-l^2}{2d_{\rm nm}R_{\rm mn}(R_{\rm mm},\theta_1)})$ \\ \hline
				$\theta_{d}(r,R_{\rm mm},\theta_1)$ & $\arccos(\frac{d_{\rm nm}^2+r^2-R_{\rm mm}^2}{2d_{\rm nm}r})$\\ \hline
				$\theta_{4}(l,R_{\rm mm},\theta_1)$ & $\pi-\theta_3(R_{\rm mm},\theta_1)-\theta_1-\arccos(\frac{l}{2R_{\rm mm}})$ \\ \hline
				$\theta_5(l,R_{\rm mm},\theta_1)$ & $\pi-\theta_3(R_{\rm mm},\theta_1)-\theta_1+\arccos(\frac{l}{2R_{\rm mm}})$ \\ \hline
				$R_{2d}(l,R_{\rm mm},\theta_1)$ & $\sqrt{(l\cos(\theta_{4})-R_{\rm mn}(R_{\rm mm},\theta_1))^2+(l\sin(\theta_4))^2}$\\\hline
				$R_{2d2}(l,R_{\rm mm},\theta_1)$ & $\sqrt{(l\cos(\theta_{5})-R_{\rm mn}(R_{\rm mm},\theta_1))^2+(l\sin(\theta_5))^2}$\\
				\hline\hline
		\end{tabular}}
	\end{center}
\end{table}
\subsection{Proof of the Distribution of Traveling Distance L}\label{proof_dis_L}
In this appendix, we provide proofs two cases: (i) $\mathbb{P}(\mathcal{N}(A_0) = 0|R_{\rm mm},\theta_1)$ and (ii) one case in $F_{\rm L}(l)$, which is condtioned on  $\theta_1+\theta_3(R_{\rm mm},\theta_1) < \pi/2$ and $2R_{\rm mm}\sin(\frac{\theta_1}{2})\leq l\leq \min(2R_{\rm mm}\sin(\theta_1),R_{\rm mn})$.

\subsubsection{Proof of $\mathbb{P}(\mathcal{N}(A_0) = 0|R_{\rm mm},\theta_1)$}

\begin{figure}[ht]
	\centering
	\includegraphics[width=0.6\columnwidth]{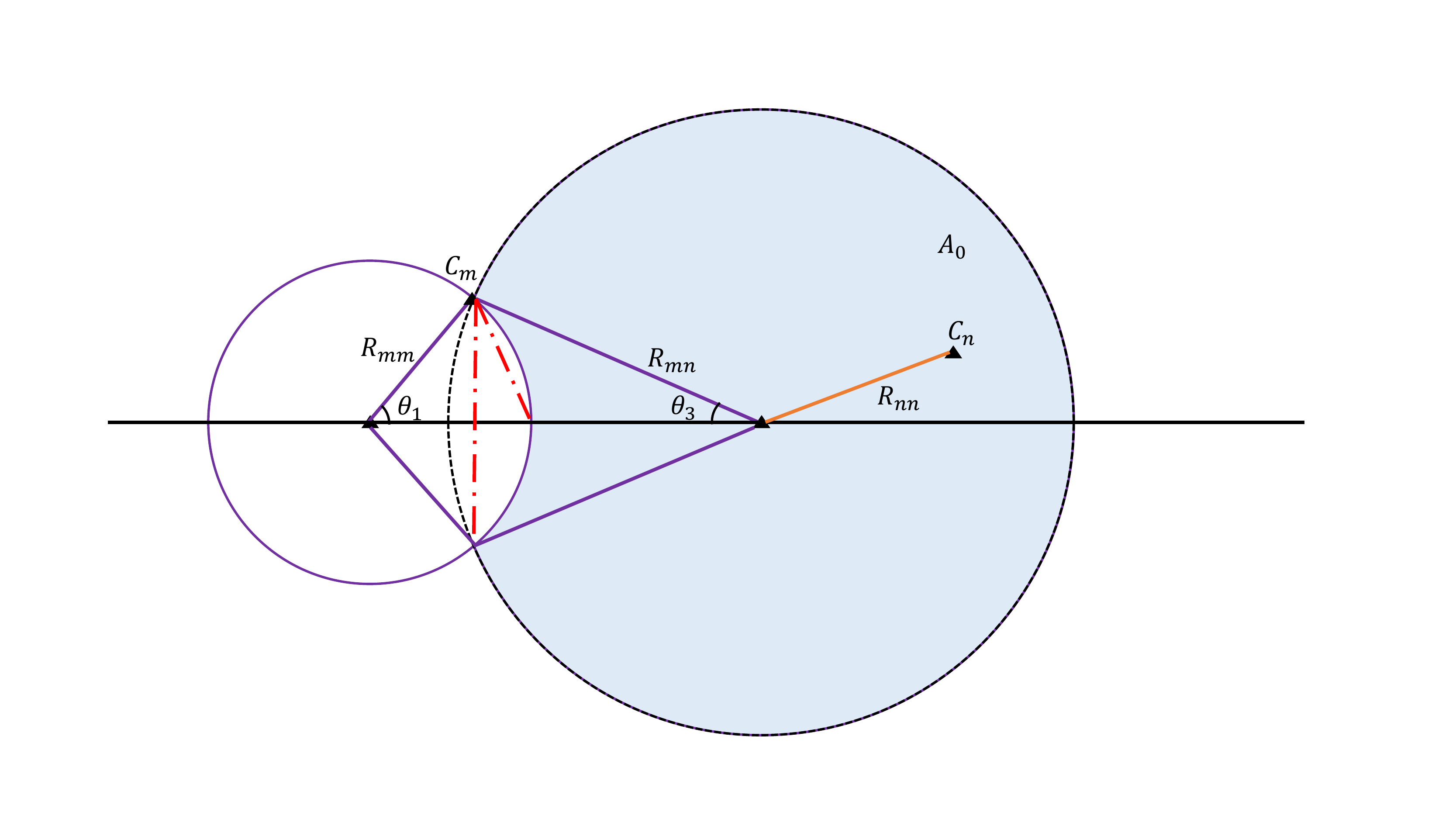}
	\caption{The possible location of $C_n$, such that $L$ is greater than 0.}
	\label{P_0}
\end{figure}
Note that in our system, we assume that $L$ is the distance between the two charging pads. Therefore, if $C_m$ locates at $R_{\rm mm}$ away and corresponding angle is $\theta_1$, $C_n$ can only be located in ${\rm A_0}$, which is colored in Fig. \ref{P_0}. Hence, the probability of $L=0$ is the probability of no point falling in $A_0$,
\begin{align}
	\mathbb{P}_{\rm (0|R_{\rm mm},\theta_1)} &= \mathbb{P}(\mathcal{N}(A_0) = 0|R_{\rm mm},\theta_1)
\end{align}
obviously, the area of $A_0$ is given by 
\begin{align}
	&|A_0|= \pi R_{\rm mm}^2 \frac{\pi-\theta_1}{\pi}+\pi R_{\rm mn}^2\frac{\pi-\theta_3}{\pi} + R_{\rm mm}d_{\rm nm}\sin\theta_1-\pi R_{\rm mm}^2,
\end{align}
$\theta_3$ and $R_{\rm mn}$ can be represented by $\theta_1$ and $R_{\rm mm}$ as,
\begin{align}
	R_{\rm mn} &= {\rm abs}\bigg(\frac{d_{\rm nm}-R_{\rm mm}\cos\theta_1}{\cos\theta_3}\bigg),  \nonumber\\
	\theta_3 &= \arccos(\frac{d_{\rm nm}^2+R_{\rm mn}^2-R_{\rm mm}^2}{2d_{\rm nm}R_{\rm mn}(R_{\rm mm},\theta_1)}), \nonumber
\end{align}
therefore, $\mathbb{P}_{\rm (0|R_{\rm mm},\theta_1)}$ now becomes,
\begin{align}
	\mathbb{P}_{\rm (0|R_{\rm mm},\theta_1)} &= \exp(-\lambda_c |A_0(R_{\rm mm},\theta_1)|),
\end{align}
where $|A_0|$ denotes the area of $A_0$.
\subsubsection{Proof of One Case In The Distribution of $L$}
In this part, we only prove one case which is $\theta_1+\theta_3(R_{\rm mm},\theta_1) < \pi/2$, $2R_{\rm mm}\sin(\frac{\theta_1}{2})\leq l\leq \min(2R_{\rm mm}\sin(\theta_1),R_{\rm mn})$ and $2R_{\rm mm}\sin(\theta_1)<R_{\rm mn}$, as shown in Fig. \ref{P_0}. Two red dash lines denote $2R_{\rm mm}\sin(\frac{\theta_1}{2})$ and $2R_{\rm mm}\sin(\theta_1)$, respectively. 

The probability of $L<l$ is the intersection of the red ball $B(C_m,l)$ and $A_0$,
\begin{align}
	F_{L}(l) &= \mathbb{P}(L<l)= 1-\mathbb{P}(L>l),
\end{align}
where $\mathbb{P}(L>l)$ is the probability of $C_n$ falling in $A_1$.

\begin{figure}[ht]
	\centering
	\subfigure{\includegraphics[width=0.49\columnwidth]{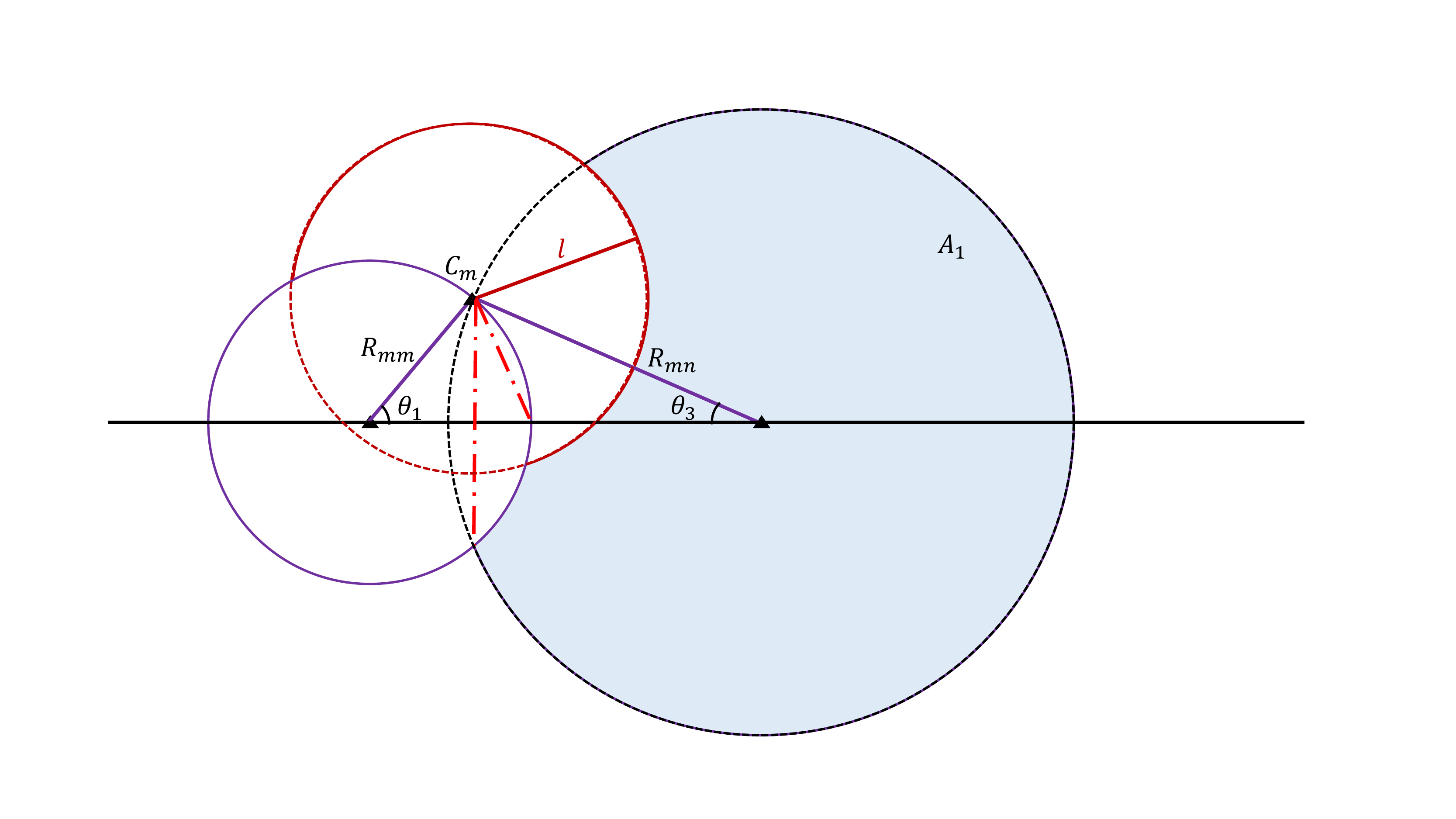}}
	\subfigure{\includegraphics[width=0.49\columnwidth]{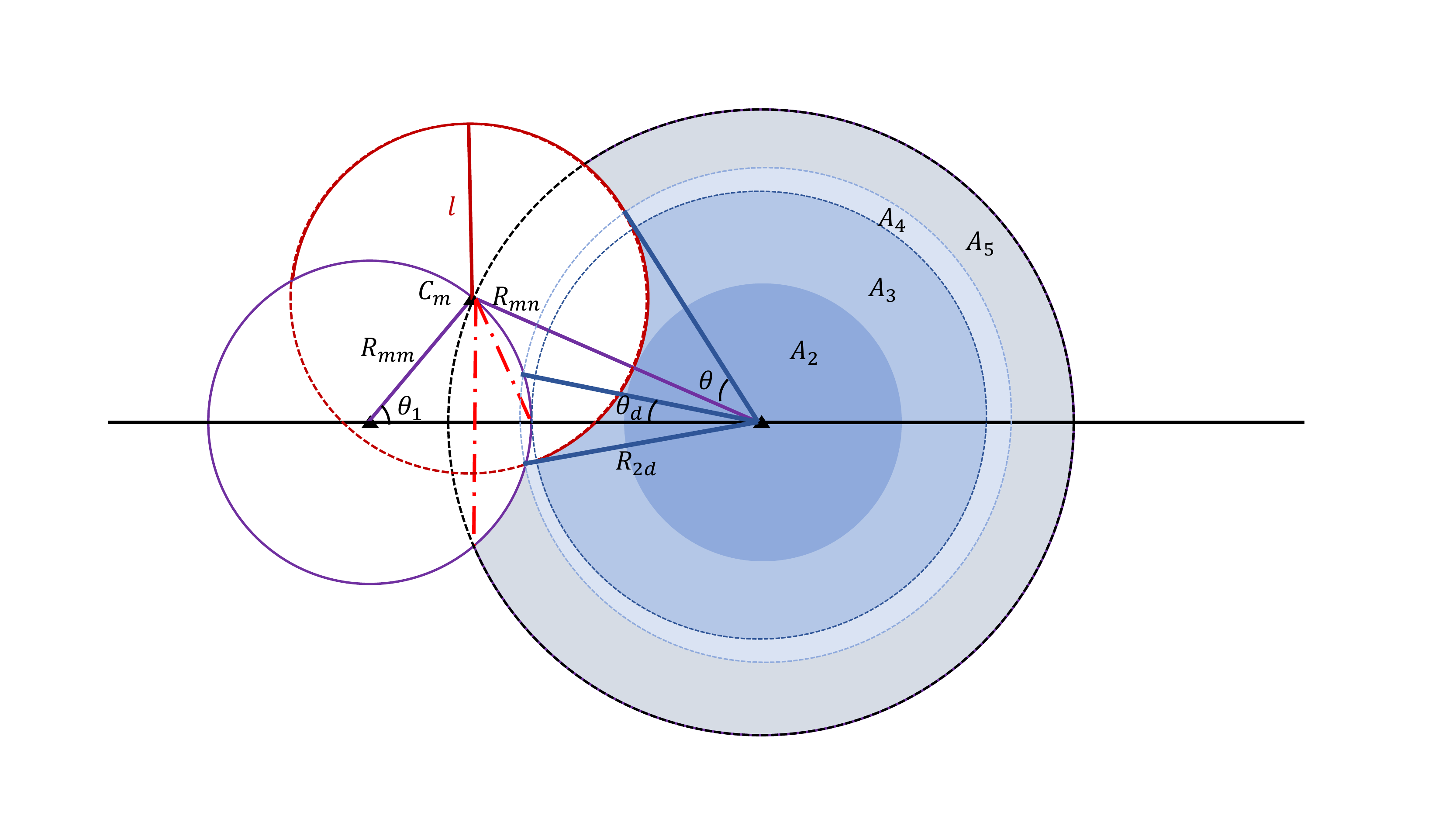}}
	\caption{Area of L not equal to 0, $\mathbb{P}_{\rm (0|R_{\rm mm},\theta_1)}$.}
	\label{4area}
\end{figure}
We then compute the probability of $C_n$ is located in $A_1$. Note that $A_1$ is consisted of $4$ areas, shown in Fig.~\ref{4area}. The probability of $C_n$ located in these 4 areas are different, we need to compute them seperately,
\begin{align}
	\mathbb{P}(L>l) &= \mathbb{P}(A_2)+\mathbb{P}(A_3)+\mathbb{P}(A_4)+\mathbb{P}(A_5),\\
	\mathbb{P}(A_2) & = 1-e^{-\lambda_{\rm c}\pi(R_{\rm mn}-l)^2},\nonumber\\
	\mathbb{P}(A_3) & = \int_{R_{\rm mn}-l}^{d_{\rm nm}-R_{\rm mm}}f_{\rm R_{\rm nn}}(r|R_{\rm mm},\theta_1)\frac{\pi-\theta(r,l,R_{\rm mm},\theta_1)}{\pi}{\rm d}r,\nonumber\\
	\mathbb{P}(A_4) & = \int_{d_{\rm nm}-R_{\rm mm}}^{R_{2d}}f_{\rm R_{\rm nn}}(r|R_{\rm mm},\theta_1)\frac{\pi-\theta(r,l,R_{\rm mm},\theta_1)}{\pi-\theta_{d}(r,l,R_{\rm mm},\theta_1)}{\rm d}r,\nonumber\\
	\mathbb{P}(A_5) & = \int_{R_{2d}}^{R_{\rm mn}}f_{\rm R_{\rm nn}}(r|R_{\rm mm},\theta_1)\frac{2\pi-\theta(r,l,R_{\rm mm},\theta_1)-\theta_3(R_{\rm mm},\theta_1)-\theta_{d}(r,R_{\rm mm},\theta_1)}{2\pi-2\theta_{d}(r,R_{\rm mm},\theta_1)}{\rm d}r,
\end{align}
where $\mathbb{P}(A_2)$ is simply derived by 1 minus the probability of no point falling in $A_1$.  Taking $C_n$ falls in $A_3$ for example, the probability of $C_n$ located at $r$ away is
\begin{align}
	\mathbb{P}(R_{\rm nn}=r) &= 2\lambda_{\rm c}r\exp(-r\pi^2)\frac{2\pi-2\theta}{2\pi}= f_{\rm R_{nn}}(r)\frac{\pi-\theta}{\pi},
\end{align}
where $f_{\rm R_{nn}}(r)$ is the distance distribuiton of $R_{\rm nn}$, $\frac{\pi-\theta}{\pi}$ denotes the area that $C_n$ can be located (except the uncolored area near $A_3$), then taking the integration over $r$ from $R_{\rm mn}-l$ to $d_{\rm nm}-R_{\rm mm}$ (the borders of $A_3$) completes the proof.

\begin{figure}[ht]
	\centering
	\includegraphics[width=0.6\columnwidth]{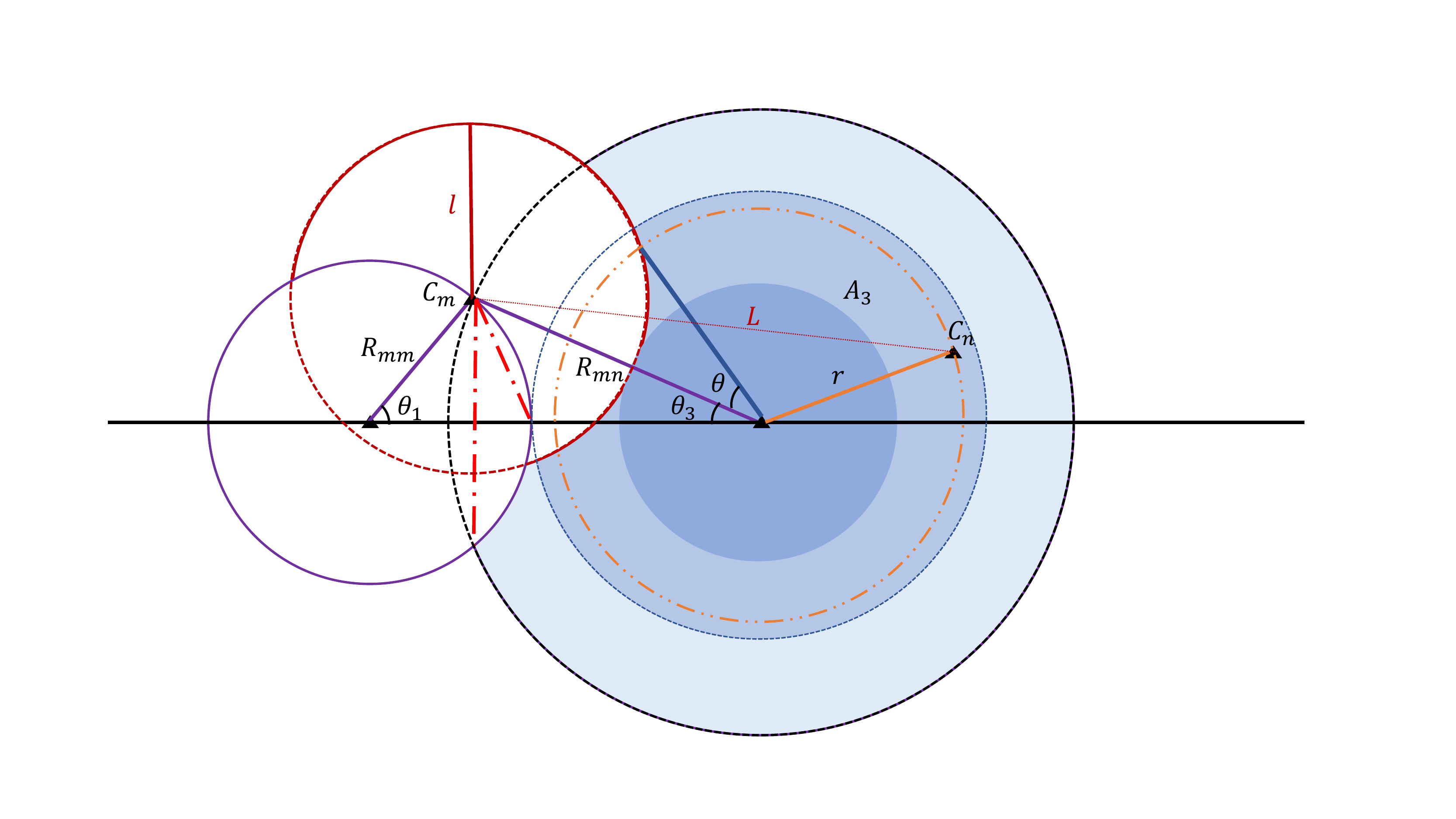}
	\caption{Area of L not equal to 0, $\mathbb{P}_{\rm (0|R_{\rm mm},\theta_1)}$.}
	\label{area3}
\end{figure}
Note that in $A_4$, the minus term $\theta_{d}$ in denominator is owing to the distance distribution of $R_{\rm nn}$, because we have already considered that the probability of $C_n$ falling in the intersection of $B(x_m,R_{\rm mm})$ and $B(x_n,R_{\rm mn})$ equals to 0.

The proof of the probabilities of $C_n$ falling in other areas follows the same method, and the geometry relationships are shown in Fig. \ref{4area}.

\subsection{Proof of Lemma \ref{lem_ass}}\label{app_ass}
If the reference user and the cluster UAV establish a LoS link with distance $r$, the association probability $\mathcal{A}_{\rm LoS}(r)$ is
\begin{align}
	\mathcal{A}_{\rm LoS}(r) &= \mathbb{P}(p_{\rm l}>p_{\rm t}) = \mathbb{P}\bigg(\eta_{\rm l}\rho_{\rm u}r^{-\alpha_{\rm l}}>\rho_{\rm t}  R_{\rm t}^{-\alpha_{\rm t}}\bigg)\nonumber\\
	&= \mathbb{P}\bigg(R_{\rm t}>(\rho_{\rm t}/\rho_{\rm u})^{\frac{1}{\alpha_{\rm t}}}\eta_{\rm l}^{-\frac{1}{\alpha_{\rm t}}}r^{\frac{\alpha_{\rm l}}{\alpha_{\rm t}}}\bigg) = \exp\bigg(-\pi \lambda_{\rm t}d_{\rm lt}^{2}(r)\bigg),
\end{align}
in which $d_{\rm lt}(r) = (\rho_{\rm t}/\rho_{\rm u})^{\frac{1}{\alpha_{\rm t}}}\eta_{\rm l}^{-\frac{1}{\alpha_{\rm t}}}r^{\frac{\alpha_{\rm l}}{\alpha_{\rm t}}}$.

Recall that $\mathcal{A}_{\rm TBS-LoS,\{m,n\}}(r,R_{\rm mm})$ denotes the event that the average received power from the nearest TBS, which locates at $r$ away, is stronger than cluster N/LoS UAV that locates at $R_{\rm um}$ away,
\begin{align}
	\mathcal{A}_{\rm TBS-LoS,m}(r,R_{\rm um}) &= \mathbbm{1}\bigg({\rm LoS(R_{\rm um})}\bigg)
	\mathbbm{1}\bigg((p_{\rm t}>p_{\rm l})\bigg)\nonumber\\
	&= \mathbbm{1}\bigg({\rm LoS(R_{\rm um})}\bigg)\mathbbm{1}\bigg(\rho_{\rm t}  r^{-\alpha_{\rm t}}>\eta_{\rm l}\rho_{\rm u}(R_{\rm um}^2+h^2)^{\frac{-\alpha_{\rm l}}{2}}\bigg)\nonumber\\
	&= \mathbbm{1}\bigg({\rm LoS(R_{\rm um})}\bigg)\mathbbm{1}\bigg(R_{\rm um}>\sqrt{d_{\rm tl}^2(r)-h^2}\bigg),
\end{align}
in which $d_{\rm tl}(r) = \max((\rho_{\rm u}/\rho_{\rm t})^{\frac{1}{\alpha_{\rm l}}}\eta_{\rm l}^{\frac{1}{\alpha_{\rm l}}}r^{\frac{\alpha_{\rm t}}{\alpha_{\rm l}}},h)$.

	\bibliographystyle{IEEEtran}
	\bibliography{ref.bib}
\end{document}